\pdfoutput=1
\documentclass[10pt]{article}
\usepackage[latin1]{inputenc}
\usepackage{amsmath}
\usepackage{amsfonts}
\usepackage{amssymb}
\usepackage{bbm}
\usepackage[all]{xy}
\usepackage[
    pdftex,
    a4paper,
    bookmarks,
    bookmarksopen=true,
    bookmarksnumbered=true,
    pdfauthor={Hein Blo"d},
    pdftitle={Meine Lebensgeschichte},
    colorlinks,
    linkcolor=blue,
    urlcolor=blue
]{hyperref}
\usepackage[footnotesize,ruled]{caption}
\usepackage[hang,nooneline]{subfigure}

\usepackage[pdftex]{graphics}
\usepackage{graphicx}

\begin{document}

\title{Orbits in the Field of a Gravitating Magnetic Monopole}

\author{Valeria Kagramanova$^1$, Jutta Kunz$^1$ and Claus L\"ammerzahl$^2$ \\
\\
$^1$
Institut f\"ur Physik, Universit\"at Oldenburg,
D--26111 Oldenburg, Germany\\
$^2$
ZARM, Universit\"at Bremen, Am Fallturm,
D--28359 Bremen, Germany}

\maketitle

\begin{abstract}
Orbits of test particles and light rays are an important tool
to study the properties of space--time metrics.
Here we systematically study the properties of the gravitational field
of a globally regular magnetic monopole in terms of
the geodesics of test particles and light.
The gravitational field depends on two dimensionless parameters,
defined as ratios of the characteristic mass scales present.
For critical values of these parameters the resulting
metric coefficients develop a singular behavior,
which has profound influence on the properties of the
resulting space-time and which is clearly reflected in the
orbits of the test particles and light rays.

\end{abstract}

PACS:

\section{Introduction}

To understand the properties of
classical solutions of the gravitational field equations
it is essential to study the orbits of test particles
and light rays in these space-times.
On the one hand,
this is important from an observational point of view, since it is only matter and light that is observed and that therefore can give insight into a given gravitational field \cite{Ehlers06}.
The study of the motion of test particles in gravitational fields
has thus significant practical applications. On the other hand, this study is also important from a fundamental point of view, since the motion of matter and light can be used to
classify a given space-time, to decode its structure
and to highlight its characteristics.

{Particles and light have been used since a long time to discuss
the properties of solutions of Einstein's field equations.
All solutions of the geodesic equation in a Schwarzschild gravitational field
can be found in a seminal paper of Hagihara \cite{Hagihara31}.
With the same mathematical tools one can solve the geodesic equation
in a Reissner--Nordstr\"om space--time \cite{Chandrasekhar83}.
The analytic solutions of the geodesic equation
in a Kerr and Kerr--Newman space--time are also known
(see \cite{Chandrasekhar83} for a survey).
Analytic solutions are the starting point for approximation methods
for the description of real stellar, planetary, comet, asteroid,
or satellite trajectories (see e.g. \cite{Hagihara70}).
Analytic solutions of the geodesic equation can also serve as test beds
for numerical codes for the dynamics of binary systems
in the extreme stellar mass ratio case
and also for the calculation of corresponding gravitational wave templates.}


Here we discuss orbits in a static spherically symmetric gravitational field,
where the space--time metric is expressed in terms of
Schwarzschild--like coordinates,
\begin{equation}
ds^2 = g_{tt} dt^2 + g_{rr} dr^2
 + r^2 (d\vartheta^2 + \sin^2\vartheta d\varphi^2) \, .
\label{metric} \end{equation}
We further assume, that the space--time is asymptotically flat.
Well-known examples of such space--times are
the Schwarzschild solution and the Reissner-Nordstr\"om solution,
both representing black hole solutions with an event horizon
and a central singularity.
In these solutions, the two metric coefficients are not independent,
but related to each other, $g_{tt} g_{rr}=-1$.
At the event horizon $g_{rr} \rightarrow \infty$,
while the curvature invariants remain finite.
At the origin, in contrast, the curvature invariants diverge,
indicating the presence of a classical singularity.

When matter fields other than the Maxwell field are coupled to
gravity, not only black hole solutions arise but
globally regular solutions can appear as well.
Interesting examples of such globally regular solutions
of the coupled Einstein-matter field equations are
boson stars, which form when a scalar field is coupled to gravity,
or magnetic monopoles, which originate in grand unified theories.
These solutions neither possess an event horizon nor
a central singularity. Their curvature invariants are finite everywhere,
and the two metric functions no longer satisfy
the relation $g_{tt} g_{rr}=-1$, instead $g_{tt} g_{rr}=-A(r)$.

Our interest here focuses on orbits of test particles
and light rays in such space--times, and, in particular,
in space--times emerging in the presence of
globally regular magnetic monopoles \cite{mono,gmono}
as described in Section \ref{Sec:GravMonopoles}.
{Magnetic monopoles arise as topological defects in theories
which undergo spontaneous symmetry breaking.
In general, magnetic monopoles exist if the mapping of the vacuum manifold,
associated with the symmetry breaking, onto the two-sphere is non-trivial.
The existence of magnetic monopoles is consequently a generic prediction
of grand unification.
If magnetic monopoles were indeed present in the universe,
they would have a host of astrophysical and cosmological consequences.
One might even think of calculating the form of gravitational waves
created by scattering processes at such monopole solutions,
since these may contribute to a stochastic gravitational wave background
emerging in the very early universe.}

For magnetic monopoles the Einstein-matter equations depend on
two dimensionless parameters,
which represent ratios of the characteristic mass scales present
in the theory. Depending on the values of these two parameters,
the metric coefficients $g_{tt}$ and $g_{rr}$
can exhibit an interesting behaviour:
the component $g_{tt}$ then approaches zero
within a certain domain, $0 \leq r \leq r_0$,
while $g_{rr}$ tends to infinity at the special value $r_0$,
and remains finite elsewhere \cite{gmono}.
We here demonstrate, that this intriguing behavior
is clearly reflected in the particle orbits.
{(When one considers SU(2) Einstein-Yang-Mills theory
without Higgs fields, also globally regular configurations
result \cite{bm}.
These solutions are unstable
\cite{review},
however, and this makes them less attractive
to study orbits in their vicinity.)}

In Section \ref{Sec:EoM} we recall the general set of equations of motion
for test particles and light rays
and also discuss the effective potential.
We then review a number of physical quantities relevant for the
interpretation of the orbits of the test particles and light rays
and discuss the general features of singular limits of
the metric functions in Section \ref{Sec:orbits}.
In Section \ref{Sec:GravMonopoles} we recall the basic equations
and main properties of gravitating non-Abelian magnetic monopoles.
In Section \ref{Sec:numerics},
the main section of the paper, we then present
numerical results obtained for the possible types of orbits
in the gravitational field of a magnetic monopole.
We here exhibit the trajectories of test particles and light rays
together with their proper time resp.~affine parameter,
and we emphasize the peculiar features of the orbits,
which arise for {\sl almost} critical values of the parameters.
We give our conclusions in Section \ref{Sec:conclusion}.

\section{Equations of motion and effective potential}\label{Sec:EoM}

\subsection{Equations of motion}

We briefly recall the equations of motion for test particles
and light in the general static spherically symmetric metric,
Eq.~(\ref{metric}).
Because of spherical symmetry, one can restrict the motion
to the equatorial plane.
One obtains two constants of motion, the specific energy $E$
and the specific angular momentum $L$
(i.e., energy and angular momentum per unit mass for a particle),
\begin{eqnarray}
E & = & - g_{tt} \frac{dt}{d\tau} \, , \label{ConservationEnergy} \\
L & = & r^2 \frac{d\varphi}{d\tau} \, ,\label{ConservationAngularMomentum}
\end{eqnarray}
where for particles $\tau$ is the proper time,
while for light it represents an affine parameter.
Then the only dynamical equation left is
\begin{subequations}
\begin{eqnarray}
\left(\frac{dr}{d\tau}\right)^2 & = & - \left( \frac{E^2}{g_{tt} g_{rr}} + \frac{1}{g_{rr}} \left(\epsilon + \frac{L^2}{r^2}\right) \right) \, , \label{BahnExact:drds} 
\end{eqnarray}
\end{subequations}
where $\epsilon = 1$ for particles and $\epsilon = 0$ for light.
Exploiting energy and angular momentum conservation
one obtains the equations for $\varphi$ and $t$ as functions of $r$
\begin{subequations}
\begin{eqnarray}
\left(\frac{dr}{d\varphi}\right)^2 & = & - \frac{r^4}{L^2} \left(\frac{E^2}{g_{tt} g_{rr}}  + \frac{1}{g_{rr}}  \left(\epsilon + \frac{L^2}{r^2}\right)\right)
\, , \\
\left(\frac{dr}{dt}\right)^2 & = & - \frac{g_{tt}^2}{E^2} \left(\frac{E^2}{g_{tt} g_{rr}} + \frac{1}{g_{rr}} \left(\epsilon + \frac{L^2}{r^2}\right)\right) \, . \label{BahnExact:drdt}
\end{eqnarray}
\end{subequations}
Eqs.~(\ref{BahnExact:drds})-(\ref{BahnExact:drdt})
then give a complete description of the dynamics.
From Eq.~(\ref{BahnExact:drds})
it is clear that the motion is restricted to the domain
$E^2 + g_{tt} \left(\epsilon + \frac{L^2}{r^2}\right) \geq 0$.


\subsection{Effective potential}

When reasoning by analogy to the Newtonian case \cite{Hartle},
a natural definition arises for an effective potential
of the Schwarzschild space--time from the geodesic equation.
For the general asymptotically flat case
we are discussing here, no such natural definition is available.
Therefore we propose a set of conditions
which we consider to be natural properties of an effective potential,
and which in our case are sufficient to give a unique definition
for an effective potential.

We thus suggest that an effective potential $U_{\rm eff}$
for an asymptotically flat space--time
should possess the following properties:
\begin{enumerate}
\item The effective potential should determine
the equation of motion for the $r$--coordinate:
$\dfrac{d^2 r}{d\tau^2} = - \dfrac{d U_{\rm eff}}{dr}$.
\item The effective potential should obey the boundary condition
$\displaystyle \lim_{r\rightarrow\infty}U_{\rm eff}(r) = 0$.
\end{enumerate}
The first condition leads to
\begin{equation}
\frac{1}{2} \left(\frac{d r}{d\tau}\right)^2 = {\cal E} - U_{\rm eff} \, ,
\end{equation}
where ${\cal E}$ is a constant.
To fix this constant we decompose
the two metric functions $g_{tt}$ and $g_{rr}$ according to
\begin{equation}
\frac{1}{g_{tt}(r)} = - 1 - f_{tt}(r) \, , \qquad 
\frac{1}{g_{rr}(r)} =   1 + f_{rr}(r) \, ,
\label{fttfrr}
\end{equation}
where $\displaystyle \lim_{r\rightarrow\infty}f_{tt}(r) = 0$
  and $\displaystyle \lim_{r\rightarrow\infty}f_{rr}(r) = 0$,
in accordance with asymptotic flatness,
and we also exploit the second condition.
The equation for the radial motion Eq.~(\ref{BahnExact:drds})
then becomes
\begin{eqnarray}
\frac{1}{2} \left(\frac{dr}{d\tau}\right)^2 & = & \frac{1}{2} \left(1 + f_{tt}(r)\right) \left(1 + f_{rr}(r)\right) E^2 - \frac{1}{2} \left(1 + f_{rr}(r)\right) \left(\epsilon + \frac{L^2}{r^2}\right)  \nonumber\\
& = & \frac{E^2 - \epsilon}{2} + \frac{1}{2} \left(f_{tt}(r) + f_{rr}(r) + f_{tt}(r) f_{rr}(r)\right) E^2 - \frac{\epsilon}{2} f_{rr}(r) - \left(1 + f_{rr}(r)\right) \frac{L^2}{2 r^2} \label{EoMEffPot} \, .
\end{eqnarray}
Except for the first term
all other terms tend to zero for $r \rightarrow \infty$,
so that we can make the unique identification
\begin{align}
{\cal E} = \frac{E^2 - \epsilon}{2} \, , \qquad U_{\rm eff} & = - \frac{1}{2} \left(f_{tt}(r) + f_{rr}(r) + f_{tt}(r) f_{rr}(r)\right) E^2 + \frac{\epsilon}{2} f_{rr}(r) + \left(1 + f_{rr}(r)\right) \frac{L^2}{2 r^2} \nonumber\\
& = \frac{1}{2} \left(E^2 - \epsilon + \frac{E^2}{g_{tt} g_{rr}} + \frac{1}{g_{rr}} \left(\epsilon + \frac{L^2}{r^2}\right)\right)\, .
\end{align}
Thus in general
the effective potential depends on both constants of motion,
on $L$ and on $E$.

Since the right hand side of Eq.~(\ref{EoMEffPot}) is quadratic in $E$,
one can recast the equation in the form \cite{Frolov}
\begin{equation}
\frac{1}{2} \left(\frac{dr}{d\tau}\right)^2 = \frac{1}{2} \left(1 + f_{tt}(r)\right) \left(1 + f_{rr}(r)\right) \left(E^2 - U(r)\right) = - \frac{1}{2} \frac{E^2 - U(r)}{g_{tt}(r) g_{rr}(r)} \, ,
\end{equation}
and identify the quantity $U$,
\begin{equation}
U = \frac{\epsilon + \frac{L^2}{r^2}}{1 + f_{tt}(r)} = - g_{tt}(r) \left(\epsilon + \frac{L^2}{r^2}\right)
\, , \label{U} \end{equation}
where $U(r) \rightarrow \epsilon$ for $r \rightarrow \infty$.
Although $U$ is sometimes called an effective potential,
it is not an effective potential in the sense of the above criteria.
However, $U$ has the big advantage of being independent of the energy $E$,
and from
\begin{equation}
E^2 - U(r) = 0
\end{equation}
one can easily determine the values of $r$,
which mark the turning points and thus give the range of the motion.
Thus the range of the motion of particles and light is
most easily obtained from $U$ and independent of $E$.
However, the full features of the geodesics are obtained
only from $U_{\rm eff}$ and are dependent on $E$.

\section{Properties of orbits}\label{Sec:orbits}

For the subsequent discussion of the orbits in the space--time
of a magnetic monopole it is instructive to first recall
a number of physical properties relevant for the
description of the motion.
These are the proper time, the physical distance and the physical
velocity.

We then discuss the limiting features of a metric,
when the metric coefficients $g_{tt}$ and $g_{rr}$
tend to zero or diverge.
The curvature invariants here indicate the
presence of a mere coordinate singularity or of a physical singularity.
We first address these limits for the Schwarzschild
and Reissner-Nordstr\"om solutions,
where $g_{tt}g_{rr}=-1$.
Then we turn to the more general case, realized
in the space--time of a gravitating magnetic monopole,
where $g_{tt}g_{rr}=-A(r)$.
Here our statements about the limits should be
interpreted to only mean ``very small'' and ``very large''.
The treatment of the limits
in the strict mathematical sense will be discussed
in a subsequent publication \cite{KKL}.

These considerations are then used in Section \ref{Sec:numerics}
in order to obtain an interpretation of the calculated trajectories.
They may be also of relevance, when one considers
a ``splitting'' of space-time or a space--time ``without time'',
features which might occur in generalized models of gravity.


\subsection{Proper time}

An important feature of the motion of a particle
is the proper time elapsing along its trajectory.
Along a radial trajectory, e.g., the proper time is given by
\begin{equation}
\tau - \tau_0 = \int_{\tau_0}^\tau d\tau
= \int_{r_0}^r \frac{d\tau}{dr} dr = \int_{r_0}^r \left(- \frac{E^2}{g_{tt} g_{rr}} - \frac{1}{g_{rr}} \left(\epsilon + \frac{L^2}{r^2}\right)\right)^{-\frac{1}{2}} dr \, . \label{ParticleProperTime}
\end{equation}

Now if a whole region arises in a space--time,
where $g_{tt}$ is very small ($g_{tt} \rightarrow 0$),
as in the case of the
nearly critical space--times of a magnetic monopole,
then in such a region $\frac{d\tau}{dr}$ is very small as well
($\frac{d\tau}{dr} \rightarrow 0$).
This means, that a particle does not feel anything like a
singularity in such a region, but that it simply needs almost no
proper time to traverse such a region.

\subsection{Proper distance}

Another relevant physical quantity is the distance
as measured from the origin in a regular region of space--time,
\begin{equation}
l = \int_0^r \sqrt{g_{rr}} dr \, .
\label{propdis} \end{equation}

When $g_{rr} \rightarrow \infty$ at some coordinate value $r_0$,
the distance to $r_0$ becomes infinite, if $\sqrt{g_{rr}}$
is not integrable at $r_0$.
The space--time in the region $0 \le r \le r_0$ then
has infinite extent.

\subsection{Physical velocity}

The velocity $dr/dt$ in Eq.~(\ref{BahnExact:drdt}) is the coordinate
velocity and not a physical velocity.
A physical velocity is measured, e.g., by an observer at rest
in the given coordinate system.
The proper time of such an observer at $r$ is given by
$dT = - \sqrt{g_{tt}} dt$,
and the physical distance to the origin (in a regular
space--time) has been defined in Eq.~(\ref{propdis}).
The radial velocity measured by such an observer at rest then is
\begin{equation}
v = \frac{dl}{dT} = \sqrt{\frac{g_{rr}}{-g_{tt}}} \frac{dr}{dt} \, .
\end{equation}
With the geodesic equation Eq.~(\ref{BahnExact:drdt})
this yields for the measured velocity
\begin{equation}
v = \sqrt{\frac{g_{rr}}{-g_{tt}}} \frac{1}{|E|} \sqrt{\frac{-g_{tt}}{g_{rr}}} \sqrt{E^2 + g_{tt} \left(\epsilon + \frac{L^2}{r^2}\right)} = \sqrt{1 + \frac{g_{tt}}{E^2} \left(\epsilon + \frac{L^2}{r^2}\right)} \, .
\label{physvelo} \end{equation}

The measured velocity
is well defined in the allowed regions of the motion,
which are restricted by the requirement
of positivity of the radicand. 
Clearly,
the measured velocity of a particle is always limited by the velocity of light.
For $\epsilon = 0$ the measured velocity is the light velocity.
We note,
that the velocity measured by an observer at rest does not depend on $g_{rr}$.

\subsection{Curvature invariants}

To differentiate between mere coordinate singularities
and physical singularities it is instructive to
consider the curvature invariants.
The simplest of these invariants is the
Kretschmann scalar $K=R_{\mu\nu\rho\sigma} R^{\mu\nu\rho\sigma}$.
For the static spherically symmetric metric Eq.~(\ref{metric})
the Kretschmann scalar is given by
\begin{eqnarray}
K
& = & \frac{4}{r^4} \left(1 - \frac{1}{g_{rr}}\right)^2 + \frac{1}{4 g_{rr}^2} \left(2 \frac{g_{tt}^{\prime\prime}}{g_{tt}} - \left(\frac{g_{tt}^\prime}{g_{tt}}\right)^2\right)^2 + \frac{2}{r^2 g_{rr}^2} \left(\frac{g_{tt}^\prime}{g_{tt}}\right)^2  \nonumber\\
& & + \frac{1}{2 g_{rr}^2} \frac{g_{rr}^\prime}{g_{rr}} \left(\frac{g_{tt}^\prime}{g_{tt}}\right)^3 - \frac{1}{g_{rr}^2} \frac{g_{tt}^{\prime\prime}}{g_{tt}} \frac{g_{tt}^\prime}{g_{tt}} \frac{g_{rr}^\prime}{g_{rr}} + \frac{1}{4 g_{rr}^2} \left(\frac{g_{tt}^\prime}{g_{tt}}\right)^2 \left(\frac{g_{rr}^\prime}{g_{rr}}\right)^2 + \frac{2}{r^2 g_{rr}^2} \left(\frac{g_{rr}^\prime}{g_{rr}}\right)^2 \, .
\end{eqnarray}

In the Schwarzschild and Reissner-Nordstr\"om solutions, for instance,
the Kretschmann scalar remains finite at the horizons,
whereas it diverges at the classical
physical singularities at the origin.



\subsection{Equations of motion for {\sl almost} singular metric coefficients}

The equations of motion Eqs.~(\ref{BahnExact:drds})-(\ref{BahnExact:drdt})
depend on the metric coefficients.
The equations for $dr/d\varphi$ and $dr/d\tau$ depend on
the metric coefficient $g_{rr}$, and on the product $g_{tt} g_{rr}$,
while the equation for $dr/dt$ further contains the factor $g_{tt}$.

To address the effects of {\sl almost} singular metric coefficients
on the orbits of particles and light,
we here distinguish two cases.
In the first case, we consider space--times,
where the metric coefficients satisfy
the relation $g_{tt} g_{rr} = - 1$,
such as, e.g., the black hole space--times
of the Schwarzschild metric and the Reissner-Nordstr\"om metric.

In the second case we consider globally regular
asympototically flat metrics, where the metric coefficients
are constrained by the above relation only at spatial infinity,
thus $g_{tt} g_{rr} = - A(r)$
with ${\rm lim}_{r \rightarrow \infty} A(r) = 1$.
These metrics include the generic
space--time of a magnetic monopole (see section
\ref{Sec:GravMonopoles}).
When certain critical values of the coupling constants are approached,
such a globally regular space--time may, however, evolve towards
a space--time with singular metric coefficients,
possibly associated with physical singularities.
Studying the orbits of particles and light
is expected to yield a deeper understanding of such limiting space--times.

\boldmath
\subsubsection{$g_{tt} g_{rr} = - 1$}
\unboldmath

When $g_{tt} g_{rr} = - 1$,
as, e.g., in the Schwarzschild and Reissner--Nordstr\"om solutions,
the equations of motion reduce to
\begin{subequations}
\begin{eqnarray}
\left(\frac{dr}{d\varphi}\right)^2 & = & \frac{r^4}{L^2}
\left( {E^2} - \frac{1}{g_{rr}}  \left(\epsilon + \frac{L^2}{r^2}\right)\right) \, , \\
\left(\frac{dr}{d\tau}\right)^2 & = & 
\phantom{  \frac{r^4}{L^2} \left( \right) }
{E^2} - \frac{1}{g_{rr}} \left(\epsilon + \frac{L^2}{r^2}\right) \, , \\
\left(\frac{dr}{dt}\right)^2 & = & \frac{g_{rr}^2}{E^2} 
\left( {E^2} - \frac{1}{g_{rr}} \left(\epsilon + \frac{L^2}{r^2}\right)\right) \, .
\end{eqnarray}
\end{subequations}

We now consider a number of effects relevant for the orbits,
which arise when $g_{rr}$ becomes very small or very large,
as collected in the following Table:
\begin{center}
\renewcommand{\arraystretch}{2.5}
\begin{tabular}{c|cccc}
 & $g_{rr} \rightarrow 0$ & $g_{rr} \rightarrow \infty$ \\ \hline
$\dfrac{dr}{d\varphi}$ & needs $E \rightarrow \infty$ & $E^2 \dfrac{r^4}{ L^2} $ \\
$\dfrac{dr}{ds}$ & needs $E \rightarrow \infty$ & 
${E^2}$ \\
$\dfrac{dr}{dt}$ & \txt{$\rightarrow \infty$ \\ (provided $E$ is large enough)} & $\rightarrow 0$ \\
$d\tau$ & 0 & finite/infinite$^*$ \\
$v$ & \txt{$\sqrt{1 - \dfrac{1}{E^2 g_{rr}} \left(\epsilon + \dfrac{L^2}{r^2}\right)}$ \\ (provided $E$ is large enough)} & $\sqrt{1 - \dfrac{1}{E^2 g_{rr}} \left(\epsilon + \dfrac{L^2}{r^2}\right)}$ \\
\end{tabular}
\renewcommand{\arraystretch}{1}
\end{center}
$^*$ depending on whether Eq.~(\ref{ParticleProperTime}) is integrable or not

\bigskip

When $g_{rr} \rightarrow \pm \infty$ ($g_{tt} \rightarrow 0$)
at some coordinate value $r_0$, a horizon is encountered,
as, e.g., in the Schwarzschild and Reissner--Nordstr\"om space--times.
The coordinate $r$ then changes from space--like to time--like
or vice versa,
unless the horizon is degenerate as in the case of an extremal
Reissner--Nordstr\"om black hole.

For $g_{rr} \rightarrow 0$ ($g_{tt} \rightarrow \pm \infty$)
on the other hand,
a singularity is approached in, e.g.,
the Schwarzschild and Reissner--Nordstr\"om space--times.
When $g_{rr} > 0$, motion is allowed only in the region close
to the singularity, if the energy is large enough
to compensate for the large negative second term
in the equations (proportional to $1/g_{rr}$).
The vanishing of $d \tau$ (or the finiteness of $\tau$)
indicates a finite proper time along a geodesic,
which is a standard criterion for the occurrence of a singularity \cite{MTW73}.


\boldmath
\subsubsection{$g_{tt} g_{rr} = - A(r)$}\label{subsec:independentgttgrr}
\unboldmath

This case is formulated to be applicable to the motion
of particles and light in the gravitational field
of a magnetic monopole, to be discussed below.
The two metric functions
are now constrained by the relation $g_{tt} g_{rr} = - 1$
only at spatial infinity, while elsewhere their product
corresponds to an independent function,
$g_{tt} g_{rr} = - A(r)$,
determined by the Einstein-matter equations.

We now consider the effects relevant for the orbits,
which arise when $g_{tt}$ and/or $g_{rr}$ become very small or very large:
\begin{center}
\renewcommand{\arraystretch}{2.5}
\begin{tabular}{c|cccc}
 & $g_{tt} \rightarrow 0$ & $g_{tt} \rightarrow -\infty$ & $g_{rr} \rightarrow 0$ & $g_{rr} \rightarrow \infty$ \\ \hline
$\dfrac{dr}{d\varphi}$ & $\rightarrow \infty$ & $\rightarrow - \dfrac{r^4}{L^2 g_{rr}} \left(\epsilon + \dfrac{L^2}{r^2}\right)$ & $\rightarrow \infty$ & $\rightarrow 0$ \\
$\dfrac{dr}{d\tau}$ & $\rightarrow \infty$ & $\rightarrow - \dfrac{1}{g_{rr}} \left(\epsilon + \dfrac{L^2}{r^2}\right)$ & $\rightarrow \infty$ & $\rightarrow 0$ \\
$\dfrac{dr}{dt}$ & $\rightarrow 0$ & $\rightarrow \infty$ & $\rightarrow \infty$ & $\rightarrow 0$ \\
$d\tau$ & 0 & finite & 0 & finite/infinite$^*$ \\
$v$ & 1 & requires large $E$ & $\sqrt{1 + \dfrac{g_{tt}}{E^2} \left(\epsilon + \dfrac{L^2}{r^2}\right)}$ & $\sqrt{1 + \dfrac{g_{tt}}{E^2} \left(\epsilon + \dfrac{L^2}{r^2}\right)}$ \\
\end{tabular}
\renewcommand{\arraystretch}{1}
\end{center}
$^*$ depending on whether Eq.~(\ref{ParticleProperTime}) is integrable or not

\bigskip
\begin{enumerate}
\item When $g_{tt} \rightarrow 0$, this may
\subitem a) occur at a specific coordinate value $r_0$,
and then signal the presence of a horizon, when it is
associated with $g_{rr} \rightarrow \pm \infty$.
\subitem b) occur within a whole region of space--time,
where $g_{rr}>0$.
The radial distance then changes very fast for small variations of the angle.
Therefore a particle crosses such a region on quasi radially straight lines.
\item When $g_{tt} \rightarrow - \infty$,
the traversed coordinate range is always finite,
while the coordinate velocity tends to become large.
The latter can be understood since in this case
it needs a very small $t$ to generate a large proper time.
\item When $g_{rr} \rightarrow 0$,
particles need to cover a huge coordinate range in order
to move a certain proper distance.
That means that $r$ has to become very large and, thus,
all quantities proportional to $dr$ become very large.
\item When $g_{rr} \rightarrow \infty$
at a specific coordinate value $r_0$,
\subitem a) a horizon is encountered, when at the same time
$g_{tt} \rightarrow 0$.
The coordinate velocity goes to zero, such that the particle appears
to be no longer able to move and cross the point $r_0$.
\subitem b) a singularity may arise, leading to a split
of space--time.
\end{enumerate}

\section{The gravitating non-Abelian monopole}\label{Sec:GravMonopoles}

In order to be able to discuss the motion of particles and light
in the gravitational field of a magnetic monopole,
we now briefly recall the basic equations and
the main features of the monopole space--times.

\subsection{Action}

$SU(2)$ Einstein-Yang-Mills-Higgs theory is described by the action
\begin{equation}
S =  \int \left\{ \frac{R}{16\pi G}
-\frac{1}{2} {\rm Tr}
\,\left( F_{\mu\nu}F^{\mu\nu} \right)
-\frac{1}{4} {\rm Tr}
\left(  D_\mu \Phi\, D^\mu \Phi  \right)
-\frac{\lambda}{4}
 {\rm Tr}\left[ \left(\Phi^2 - v^2 \right)^2 \right]
\right\}
\sqrt{-g} d^4 x
\ , \label{action} \end{equation}
with curvature scalar $R$,
$su(2)$ field strength tensor
\begin{equation}
F_{\mu\nu} = \partial_\mu A_\nu - \partial_\nu A_\mu + i e [A_\mu, A_\nu] \ ,
\end{equation}
gauge potential $A_\mu = A_\mu^a \tau^a/2$,
and covariant derivative of the Higgs field $\Phi = \Phi^a \tau^a$
in the adjoint representation
\begin{equation}
D_\mu \Phi = \partial_\mu \Phi +i e [A_\mu, \Phi] \ .
\end{equation}
Here $G$ and $e$ denote the gravitational and gauge coupling constants,
respectively,
$v$ denotes the vacuum expectation value of the Higgs field,
and $\lambda$ represents the strength of the Higgs self-coupling.

The nonzero vacuum expectation value of the Higgs field
breaks the non-Abelian $SU(2)$ gauge symmetry to an Abelian $U(1)$ symmetry.
The particle spectrum of the theory then consists of a massless photon,
two massive vector bosons of mass $M_W = e v$,
and a Higgs field of mass $M_H = {\sqrt {2 \lambda}}\, v$.
In the limit $\lambda \rightarrow 0$ the Higgs potential vanishes,
and the Higgs field becomes massless.

\subsection{General equations of motion}

Variation of the action Eq.~(\ref{action}) with respect to the metric
$g^{\mu\nu}$ leads to the Einstein equations
\begin{equation}
G_{\mu\nu}= R_{\mu\nu}-\frac{1}{2}g_{\mu\nu}R = 8\pi G T_{\mu\nu}
\  \label{ee} \end{equation}
with stress-energy tensor
\begin{eqnarray}
T_{\mu\nu} 
  &=&
      2\, {\rm Tr}\,
    ( F_{\mu\alpha} F_{\nu\beta} g^{\alpha\beta}
   -\frac{1}{4} g_{\mu\nu} F_{\alpha\beta} F^{\alpha\beta}) \nonumber \\
  &+&
      {\rm Tr}\, (\frac{1}{2}D_\mu \Phi D_\nu \Phi
    -\frac{1}{4} g_{\mu\nu} D_\alpha \Phi D^\alpha \Phi)
   -\frac{\lambda}{8}g_{\mu\nu} {\rm Tr}(\Phi^2 - v^2)^2
\ .
\end{eqnarray}

Variation with respect to the gauge potential $A_\mu$
and the Higgs field $\Phi$
leads to the matter field equations,
\begin{eqnarray}
& &\frac{1}{\sqrt{-g}} D_\mu(\sqrt{-g} F^{\mu\nu})
   -\frac{1}{4} i e [\Phi, D^\nu \Phi ] = 0 \ ,
\label{feqA} \\
& & \frac{1}{\sqrt{-g}} D_\mu(\sqrt{-g} D^\mu \Phi)
+\lambda (\Phi^2 -v^2) \Phi  = 0 \ .
\label{feqPhi}
\end{eqnarray}

\subsection{Ans\"atze}

To obtain a static spherically symmetric
magnetic monopole with unit magnetic charge
we parametrize the metric in Schwarz\-schild-like coordinates
\cite{mono,gmono}
\begin{equation}
ds^2=g_{tt}(r)dt^2 + g_{rr}(r)dr^2
 + r^2 (d\theta^2 + \sin^2\theta d\phi^2)
\label{ansatzg}
\ , \end{equation}
and introduce the mass function $m(r)$
\begin{equation}
g_{rr}(r) = \left( 1-\frac{2 m(r)}{r} \right)^{-1}
\label{ansatzN}
\ . \end{equation}

For the gauge potential and the Higgs field we employ the
spherically symmetric Ans\"atze \cite{mono,gmono}
\begin{equation}
A_\mu dx^\mu=
\frac{1-K(r)}{2e} \left( \tau_\varphi d\theta
                        -\tau_\theta \sin\theta d\varphi \right)
\label{ansatzA}
\ , \end{equation}
and
\begin{equation}
\Phi = v H(r) \tau_r \ ,
\label{ansatzPhi}
\end{equation}
where the $su(2)$ matrices $\tau_r$, $\tau_\theta$ and $\tau_\varphi$
are defined as products of the spherical spatial unit vectors
with the vector of Pauli matrices $\tau^a$.

\subsection{Dimensionless quantities}

We now introduce the dimensionless coupling constants
$\alpha$ and $\beta$ as ratios of the mass scales
present in the theory,
\begin{equation}
\alpha^2 = \frac{4 \pi}{g^2} \frac{M_W^2}{M_{Pl}^2} = 4 \pi G v^2 \ , \ \ \
\beta^2 = \frac{1}{2} \frac{M_H^2}{M_W^2} = \frac{\lambda}{e^2}
\ , \end{equation}
and $M_{Pl}=1/\sqrt{G}$ is the Planck mass.
We further pass to dimensionless coordinates,
$e v r \rightarrow r$,
and a dimensionless mass function,
$e v m \rightarrow m$.

\subsection{Monopole equations of motion}

The $tt$ and $rr$ components of the Einstein equations then yield
equations for the mass function $m$
and for the product $g_{tt} g_{rr}$,
\begin{eqnarray}
m'&=&\alpha^2 \Biggl(
       {K'}^2 + \frac{1}{2}   r^2 {H'}^2
     -\frac{2 m}{r} {K'}^2 - m r {H'}^2
   + \frac{(K^2-1)^2}{2 r^2} + H^2 K^2
   + \frac{\beta^2}{4} r^2 (H^2-1)^2 \Biggr)
\ , \end{eqnarray}
and
\begin{eqnarray}
 \left(g_{tt} g_{rr} \right) '&=& 2 \alpha^2 r \Biggl(
     \frac{2 {K'}^2}{r^2} + {H'}^2 \Biggr) \left(g_{tt} g_{rr} \right)
\ , \label{eqa} \end{eqnarray}
where the prime indicates the derivative with respect to $r$.
For the matter functions we obtain the equations
\begin{eqnarray}
\frac{1}{\sqrt{-g_{tt} g_{rr}}} \left(
\frac{\sqrt{ -g_{tt} g_{rr}} K' }{g_{rr}} \right)' =
 K \left( \frac{K^2-1}{r^2} + H^2
 \right)
\ , \end{eqnarray}
and
\begin{eqnarray}
\frac{1}{\sqrt{- g_{tt} g_{rr}}} \left(
\frac{ r^2 \sqrt{- g_{tt} g_{rr}} H' }{g_{rr}} \right)' =
 H \left( 2 K^2 + \beta^2 r^2 (H^2-1) \right)
\ . \end{eqnarray}
This set of equations depends only on the dimensionless
coupling constants $\alpha$ and $\beta$.
Since no analytic solutions to these field equations are known,
we use numerical solutions for the calculation of the test particle orbits.

\subsection{Boundary conditions}

To obtain globally regular particle-like solutions
we require at the origin the boundary conditions
\begin{equation}
m(0)=0 \ , \ \ \ K(0)=1 \ , \ \ \ H(0)=0
\, . \end{equation}

Asymptotic flatness implies
that the metric functions $g_{tt}$ and $g_{rr}$
approach constants at infinity.
We adopt
\begin{equation}
g_{tt}(\infty) g_{rr}(\infty)=-1 \, .  \label{BoundaryCondition}
\end{equation}
$m(\infty)$ represents the dimensionless mass
of the monopole solutions.

The matter functions satisfy asymptotically
\begin{equation}
K(\infty)=0 \ ,
\ \ \ H(\infty) = 1
\ . \end{equation}

\subsection{Embedded Reissner-Nordstr\"om solutions}

A special black hole solution of the set of coupled equations is the embedded
Reissner-Nordstr\"om solution with unit magnetic charge,
\begin{equation}
m(r) = m_\infty - \frac{\alpha^2 }{2r} , \ \ \ g_{tt}(r) g_{rr}(r) =-1
\ , \end{equation}
\begin{equation}
K(r)=0 \ , \ \ \
H(r)=1
\ . \label{RN} \end{equation}

The extremal Reissner--Nordstr\"om
solution with event horizon radius $r_{\rm H}$ satisfies
\begin{equation}
r_{\rm H} = m_\infty = \alpha
\ . \label{RNex} \end{equation}

\subsection{Properties of magnetic monopole solutions}

The coupling to gravity has a significant effect on the magnetic
monopole solutions present in flat space \cite{mono}.
When the coupling constant $\alpha$ is increased from zero,
a branch of gravitating monopole solutions emerges smoothly
from the flat space 't Hooft-Polyakov monopole solution.
This branch of gravitating monopole solutions
extends up to a maximal value $\alpha_{\rm max}$,
beyond which gravity becomes too strong
for regular monopole solutions to persist \cite{gmono}.

For vanishing coupling constant $\beta$,
this first gravitating monopole branch merges with a second
branch at $\alpha_{\rm max}$,
which extends slightly backwards,
up to a critical value $\alpha_{\rm cr}$ of the coupling constant.
At $\alpha_{\rm cr}$ this second branch of
gravitating monopole solutions bifurcates with the branch of
extremal Reissner--Nordstr\"om solutions.

In particular, at $r_0=\alpha_{\rm cr}$ a
double zero of the metric function $1/g_{rr}$ appears.
But $r_0$ does not correspond to a degenerate horizon.
The Kretschmann scalar diverges there.
The exterior space--time of the solution, however,
corresponds to the one of an extremal Reissner--Nordstr\"om
black hole with unit magnetic charge \cite{gmono}.
The interior space--time retains regularity at the center,
due to the influence of the non-Abelian fields present.

As $\beta$ increases, the second branch decreases in size,
until at a certain value of $\beta$ the maximal value $\alpha_{\rm max}$
and the critical value $\alpha_{\rm cr}$ coincide \cite{gmono}.
For a considerably larger value of $\beta$
another interesting phenomenon arises.
The metric function $1/g_{rr}$ develops a second minimum,
and the double zero now arises at a value $r^*< r_0$,
where $r^*$ does correspond to a degenerate horizon.
The critical solution then corresponds
to an extremal black hole with non-Abelian hair
and with a mass less than that of the corresponding
extremal Reissner--Nordstr\"om solution \cite{lue}.

\section{Motion in the space--time of a gravitating monopole}\label{Sec:numerics}

This section constitutes the central part of the work.
Here we present and discuss the possible types of orbits
in the space--time of a gravitating monopole.
The monopole solutions depend on the dimensionless parameters
$\alpha$ and $\beta$. The parameters then enter via the
respective metric functions in the equations of motion
and determine via the potentials the allowed trajectories
of particles and light.

We first address the potential $U$ and the effective potential
$U_{\rm eff}$ in these monopole space--times,
considering, in particular, their expansion close to the origin.
We then discuss the orbits of particles and light
obtained numerically in these space--times.
Here we exhibit generic examples of the possible types of motion,
and we discuss motion in space--times
very close to critical monopole solutions.

One of the interesting features of the motion
in the space--time of a monopole
is the capture of light rays by the source when $\alpha$ is big:
here light is found to move on bound geodesics.
Large $\alpha$ space--times also admit two bound regions for particles
to move in.

\subsection{Potentials}

The space--time of a generic non-Abelian magnetic monopole
is globally regular and asymptotically flat.
Only when critical values of the parameters are approached
the space--time may evolve coordinate or physical
singularities.

Asymptotic flatness implies appropriate boundary conditions
for the metric functions, which are reflected in the
fall-off of the functions $f_{rr}$ and $f_{tt}$,
introduced in Eq.~(\ref{fttfrr}) to define the effective potential.
From Eq.~(\ref{ansatzN}) we identify $f_{rr}$ for the monopole
space--time,
\begin{equation}
f_{rr}(r) = - \frac{2 m(r)}{r}
\end{equation}
which vanishes for $r \rightarrow \infty$.
From the monopole boundary condition Eq.~(\ref{BoundaryCondition})
we also infer that $f_{tt} \rightarrow 0$ for $r \rightarrow \infty$.
Thus the effective potential $U_{\rm eff}$ has the required
asymptotic behaviour.

The generic monopole space--time is globally regular,
i.e., there is, in particular, no singularity present at the origin,
but the space--time is smooth there.
This is seen in the expansions of the metric functions at the origin,
since for small $r$ one finds \cite{gmono}
\begin{equation}
f_{rr}(r) = - c \alpha^2 r^2 + {\cal O}(r^4) \, , \qquad
f_{tt}(r) = c^{\prime} + {\cal O}(r) \, ,
\end{equation}
where $c$ and $c^{\prime}$ are constants.
It is also reflected in the smooth behaviour of the curvature
invariants close to the origin.

Close to the origin, the potentials then have the following expansion
\begin{eqnarray}
U_{\rm eff} & = & \frac{L}{2 r^2} + d + {\cal O}(r) \, , \\
U & = & 
\frac{L^2}{2 r^2} + \frac{d^{\prime}}{r} + d^{\prime\prime} + {\cal O}(r) \, ,
\end{eqnarray}
where $d$, $d^{\prime}$, and $d^{\prime\prime}$ are constants.
Therefore, for $r \rightarrow 0$ a repulsive angular momentum barrier
is always present in the effective potential,
unless the motion is purely radial.
In orbits with finite angular momentum
particles or light rays can never reach the origin, $r = 0$.

\begin{figure}[t]
\begin{center}
\subfigure[][The potential
$U$]{\includegraphics[width=0.45\textwidth]{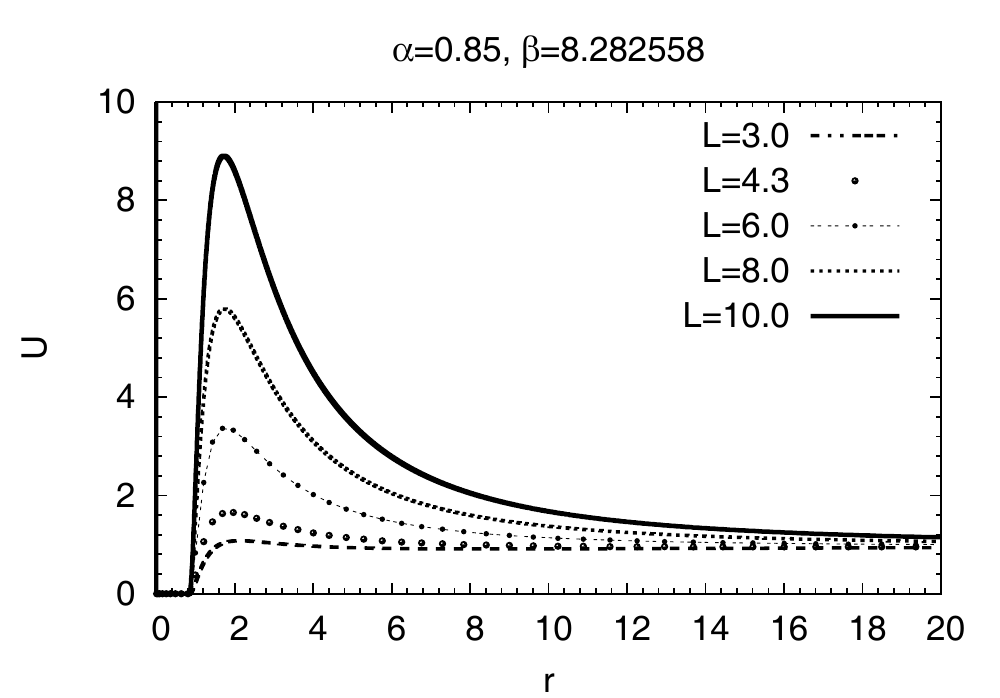}}
\subfigure[][The effective potential $U_{\rm
eff}$]{\includegraphics[width=0.45\textwidth]{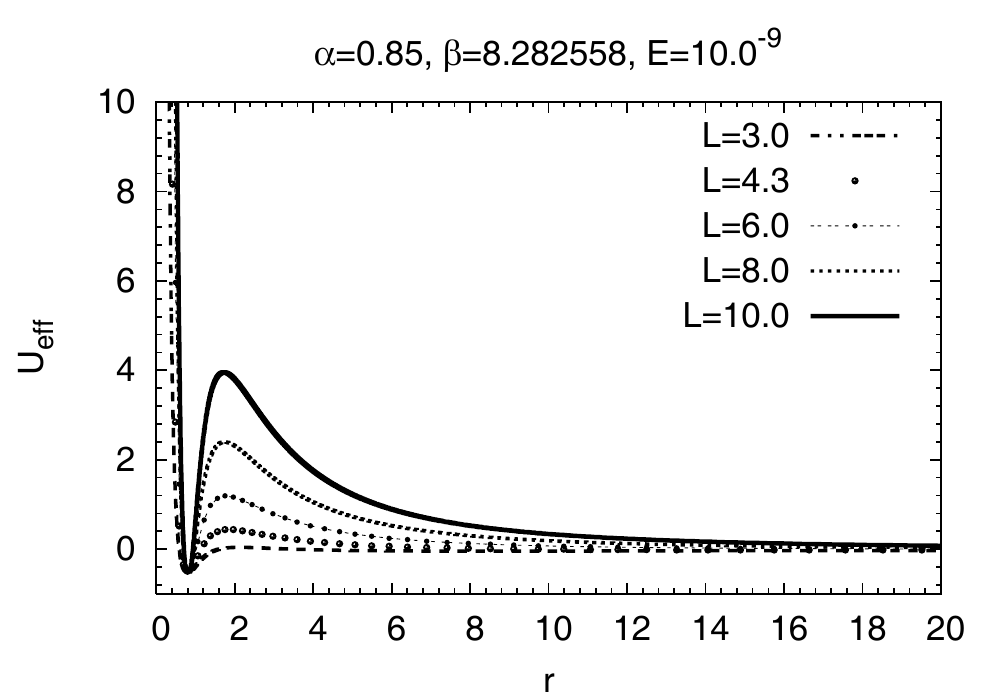}}
\end{center}
\caption{
The potential $U$ (a) and the effective potential $U_{\rm eff}$
at energy $E=10^{-9}$ (b)
versus the radial coordinate $r$
for the parameters $(\alpha, \beta)=(0.85,8.282558)$
and several angular momenta $L$ (see subsection \ref{lastsub}).
\label{Fig:AngularMomenta}
}
\end{figure}

As in the Newtonian case,
the angular momentum barrier here dominates close to origin.
But while the angular momentum barrier increases with
increasing $L$, its influence is modulated by the
metric function $g_{tt}$,
so that in regions with very small $|g_{tt}|$
the influence of the angular momentum barrier is reduced.
This is seen in Fig.~\ref{Fig:AngularMomenta},
where the potential $U$ and the effective potential $U_{\rm eff}$
are shown for $(\alpha, \beta)=(0.85,8.282558)$.


\boldmath
\subsection{Orbits for $\beta=0$}
\unboldmath

We now discuss the possible orbits for particles and light rays
in monopole space--times at vanishing $\beta$.
We begin with a set of typical solutions, obtained for
$(\alpha,\beta)=(0.25,0)$.
As $\alpha$ is increased a maximal value $\alpha_{\rm max}$ is reached,
which still allows for the existence of a globally regular monopole solution.
Beyond $\alpha_{\rm max}$ only black hole solutions exist.
We exhibit a set of typical orbits present at $\alpha_{\rm max} = 1.403$.

When $\beta=0$ (or small),
two regular monopole solutions exist
in the range $\alpha_{\rm cr} < \alpha \le \alpha_{\rm max}$.
For $\beta=0$, the critical value of $\alpha$
is determined to be (within numerical accuracy)
$\alpha_{\rm cr} = 1.385853$.
It is called critical, because as $\alpha \rightarrow \alpha_{\rm cr}$
the minimum of the metric function $1/g_{rr}$ decreases,
and reaches zero at a certain value $r_0$ of the radial coordinate
in the limit.
The region of space--time with $r\ge r_0$
of the critical solution then corresponds to the exterior space--time
of an extremal Reissner-Nordstr\"om black hole.
We exhibit a set of orbits at $\alpha_{\rm cr} = 1.385853$,
i.e., for an {\sl almost} critical space--time.

In the following we always present figures for
the metric functions $-g_{tt}$ and $1/g_{rr}$, the curvature invariant $K$,
and the potential $U$, and exhibit various typical and special orbits obtained
for given values of the angular momentum $L$ and the energy $E$.



\begin{figure}[t!]
\begin{center}
\subfigure[][$-g_{tt}$]{\includegraphics[width=0.4\textwidth]{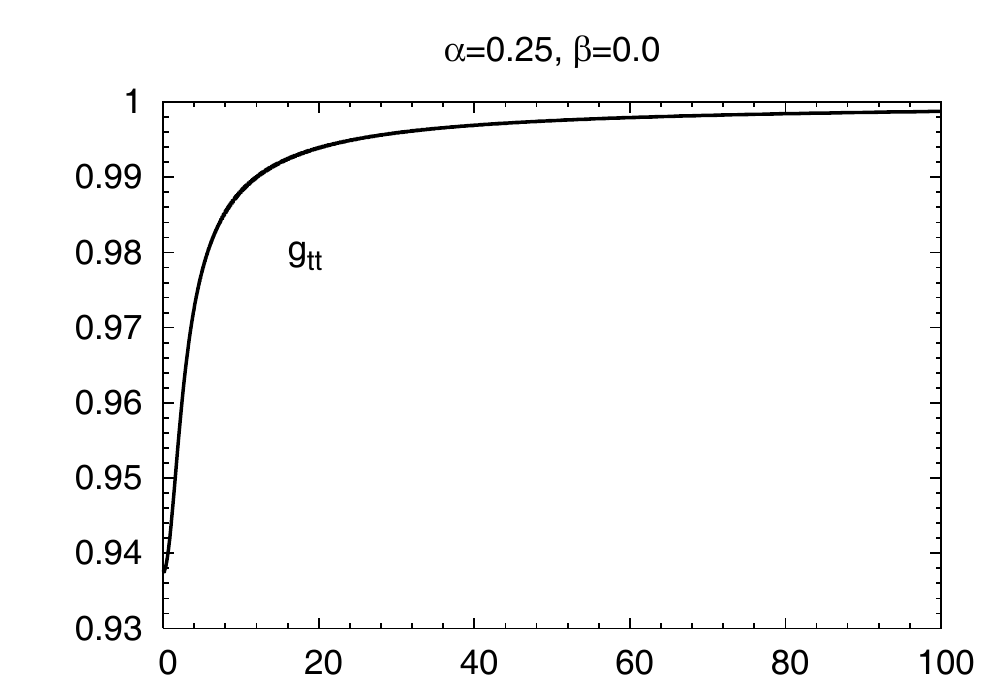}}
\subfigure[][$1/g_{rr}$]{\includegraphics[width=0.4\textwidth]{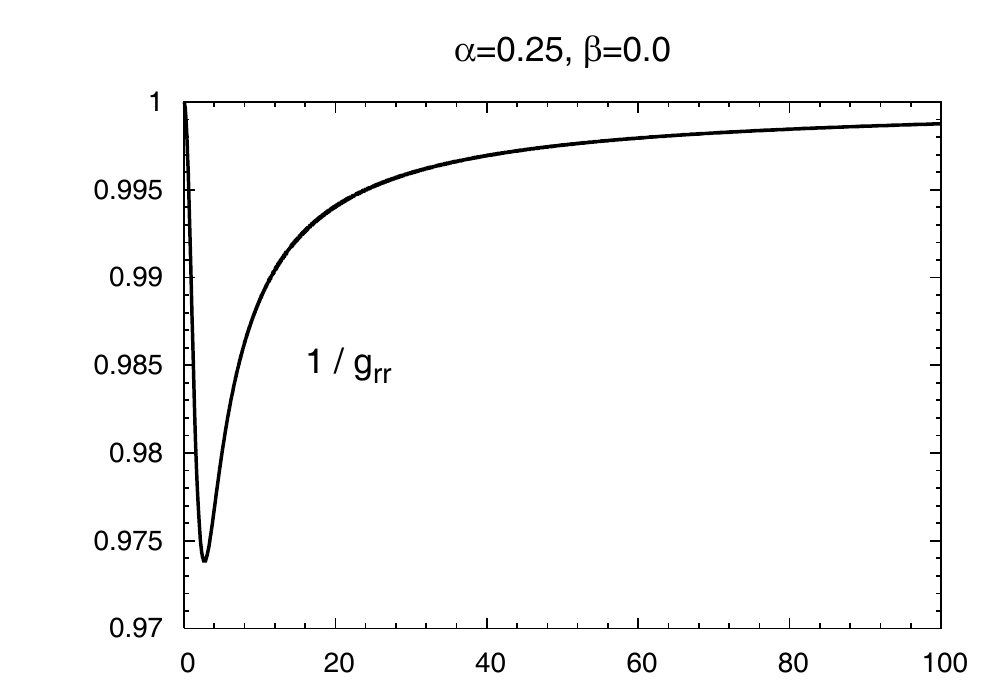}}
\subfigure[][$U$]{\includegraphics[width=0.4\textwidth]{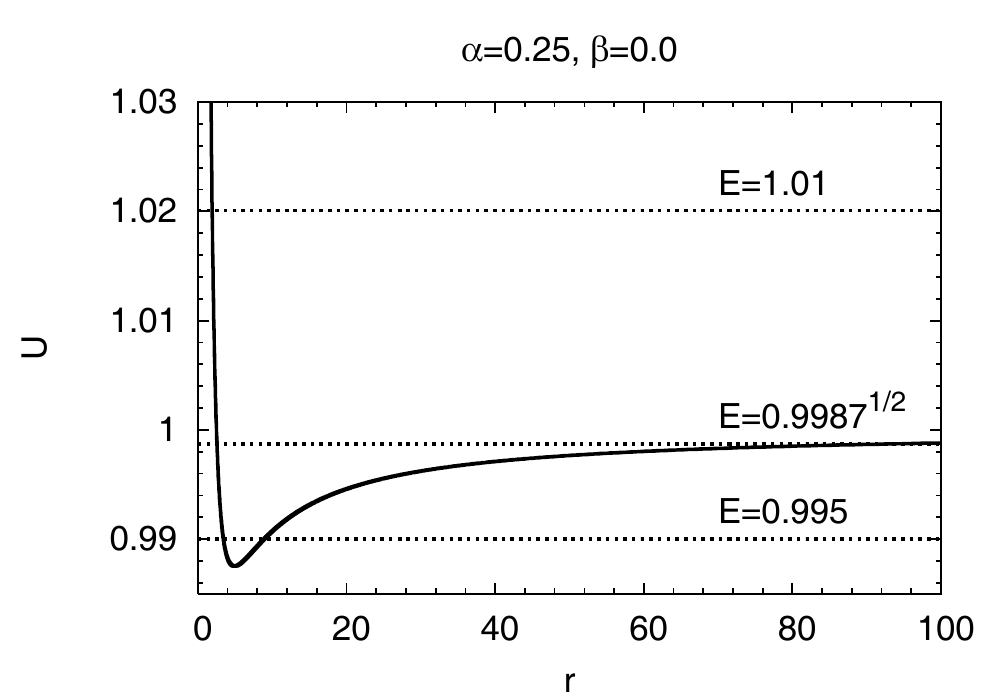}}
\subfigure[][$K$]{\includegraphics[width=0.4\textwidth]{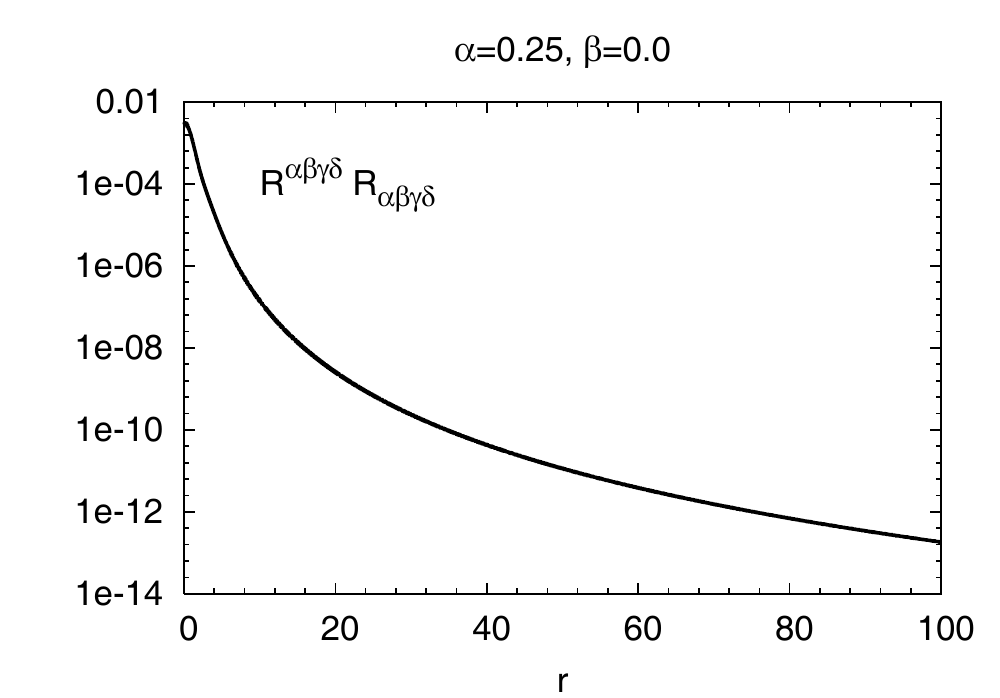}}
\end{center}
\caption{The metric functions $-g_{tt}$ (a) and $1/g_{rr}$ (b),
the potential $U$ (c) and the curvature invariant $K$ (d)
versus the radial coordinate $r$
for $(\alpha,\beta)=(0.25,0)$. \label{series1a}}
\end{figure}

\begin{figure}[h!]
\begin{center}
\subfigure[][low energy bound orbit]{\includegraphics[width=0.3\textwidth]{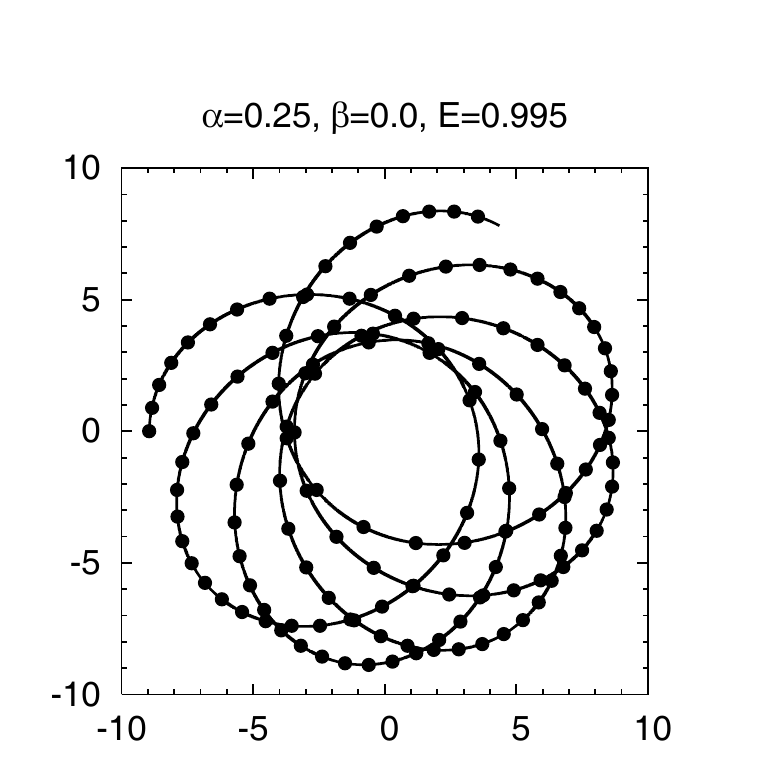}}
\subfigure[][high energy bound orbit]{\includegraphics[width=0.3\textwidth]{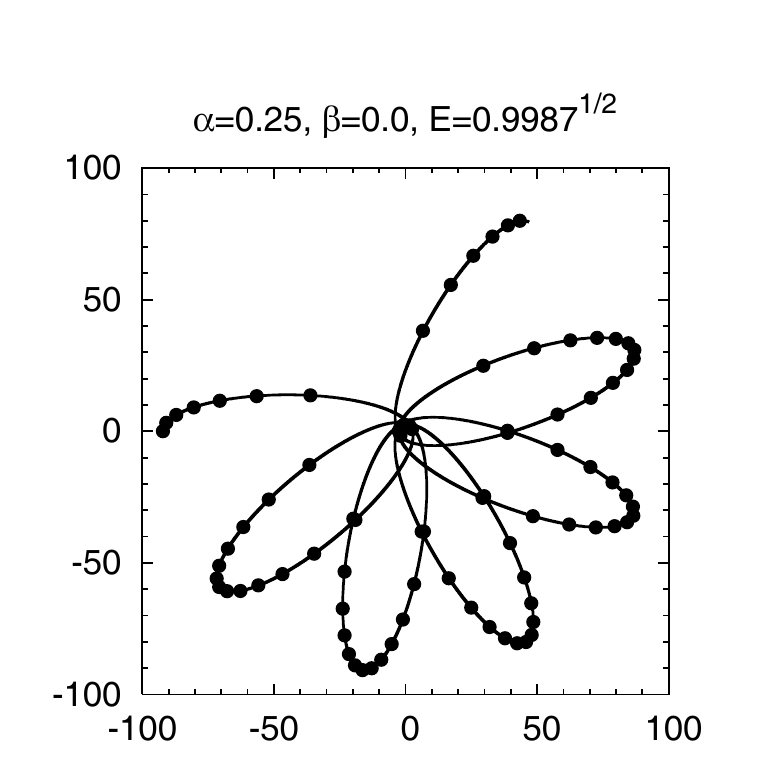}}
\subfigure[][quasi--hyperbolic orbit]{\includegraphics[width=0.3\textwidth]{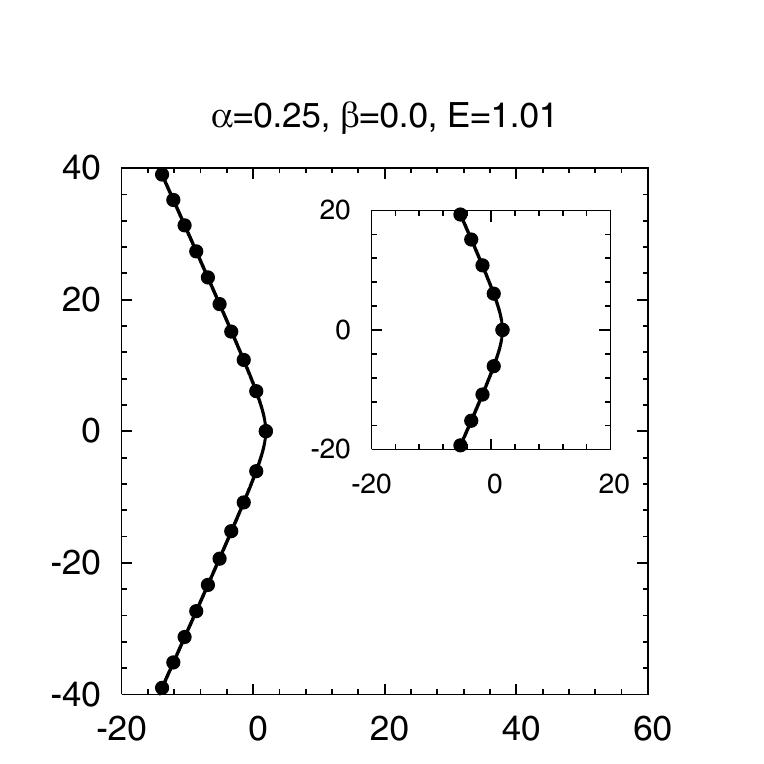}}
\end{center}
\caption{
Particle orbits $r(\varphi)$ for $(\alpha,\beta)=(0.25,0)$.
The dots indicate units of elapsed proper time.
\label{series1b}}
\end{figure}


\boldmath
\subsubsection{Orbits at $\alpha=0.25$}
\unboldmath

We begin with the discussion of the generic case of a
monopole space--time, as exhibited in Fig.~\ref{series1a}
for $(\alpha,\beta)=(0.25,0)$.
The metric coefficient $-g_{tt}$ has a finite value at the origin
and rises monotonically to its asymptotic value, $-g_{tt}(\infty) = 1$.
The metric coefficient $1/g_{rr}$ assumes the value $1/g_{rr}=1$ both at
the origin and asymptotically, and exhibits a minimum at some
value of the radial coordinate, $r_{0}$.
The Kretschmann scalar is finite everywhere and small.

The potential $U$ shows,
that the orbits in such a space--time
have the general structure of typical Newtonian orbits:
there is one circular orbit corresponding to the minimum of $U$,
and there are bound orbits as well as scattering states.

To exhibit the possible types of orbits,
we choose as a representative value for the angular momentum, $L=0.5$.
As seen in Fig.~\ref{series1b},
the orbits indeed show the qualitative structure inferred from $U$.
The main difference to the typical Newtonian orbits
is the occurrence of a perihelion (perimonopolion) shift for bound orbits,
showing that the underlying potential differs
from the Newtonian potential of a point mass.
The elapsed proper time along the orbits is also indicated.

\boldmath
\subsubsection{Orbits at $\alpha_{\rm max}=1.403$}{\label{secondsub}}
\unboldmath

\begin{figure}[h!]
\begin{center}
\subfigure[][$-g_{tt}$]{\includegraphics[width=0.4\textwidth]{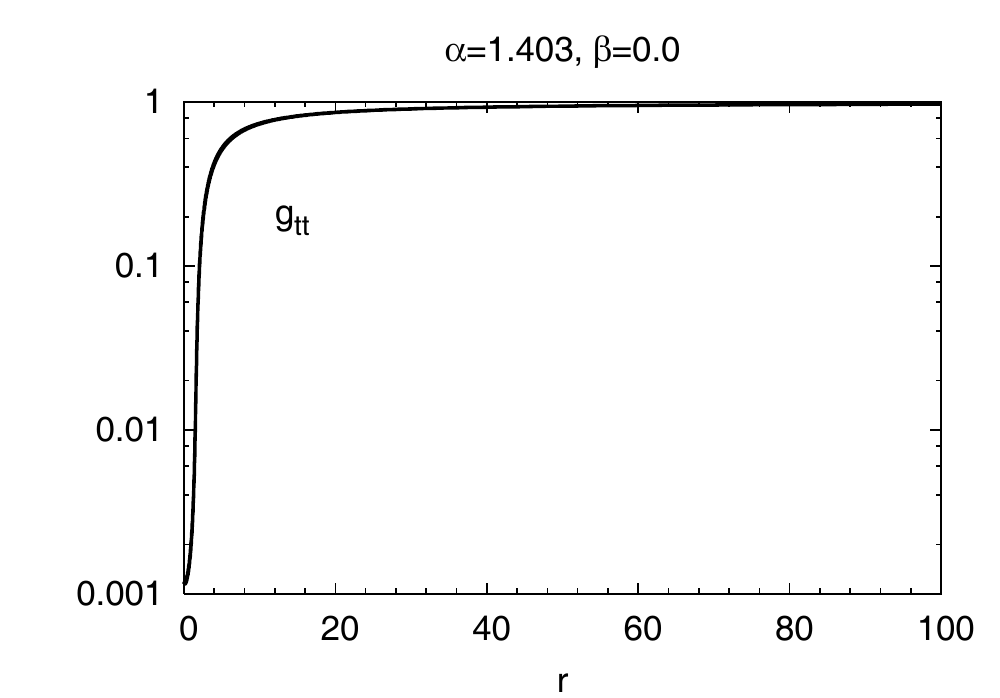}}
\subfigure[][$1/g_{rr}$]{\includegraphics[width=0.4\textwidth]{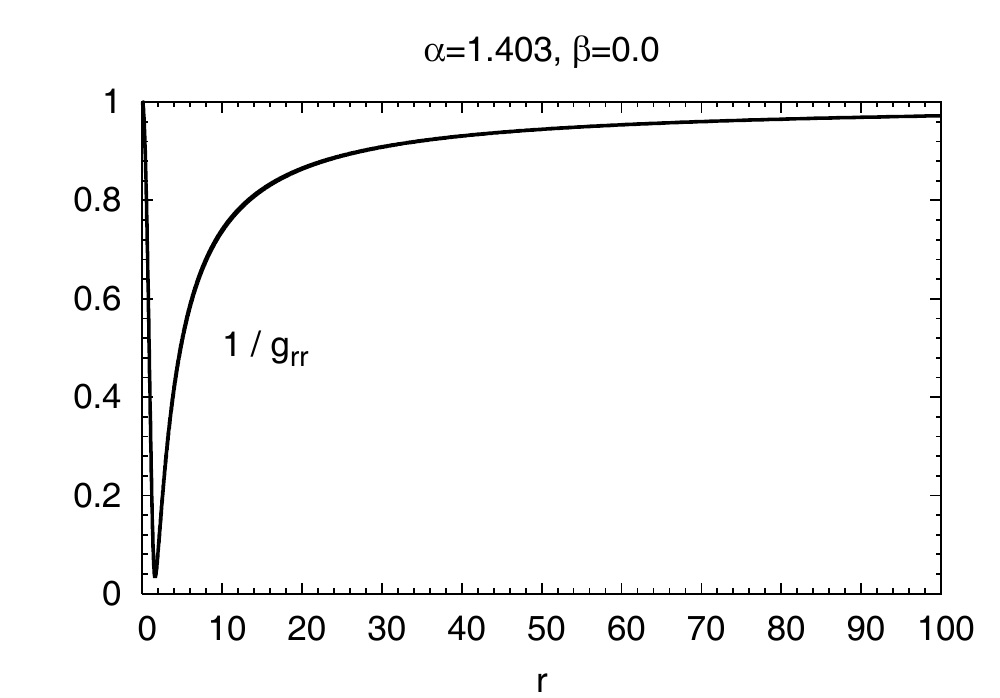}}
\subfigure[][$U$]{\includegraphics[width=0.4\textwidth]{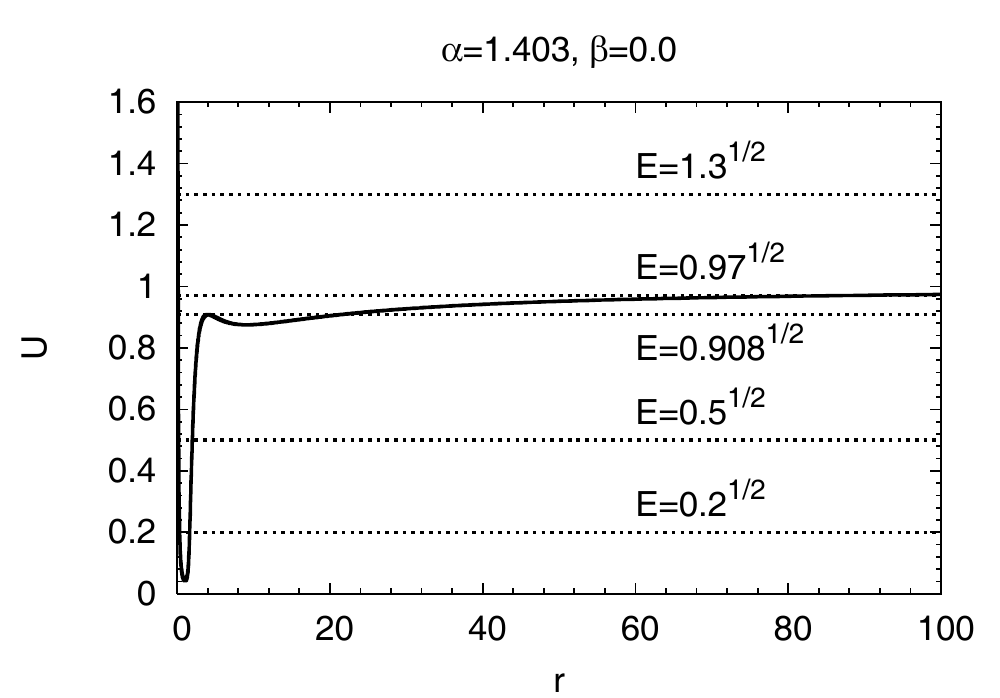}}
\subfigure[][$K$]{\includegraphics[width=0.4\textwidth]{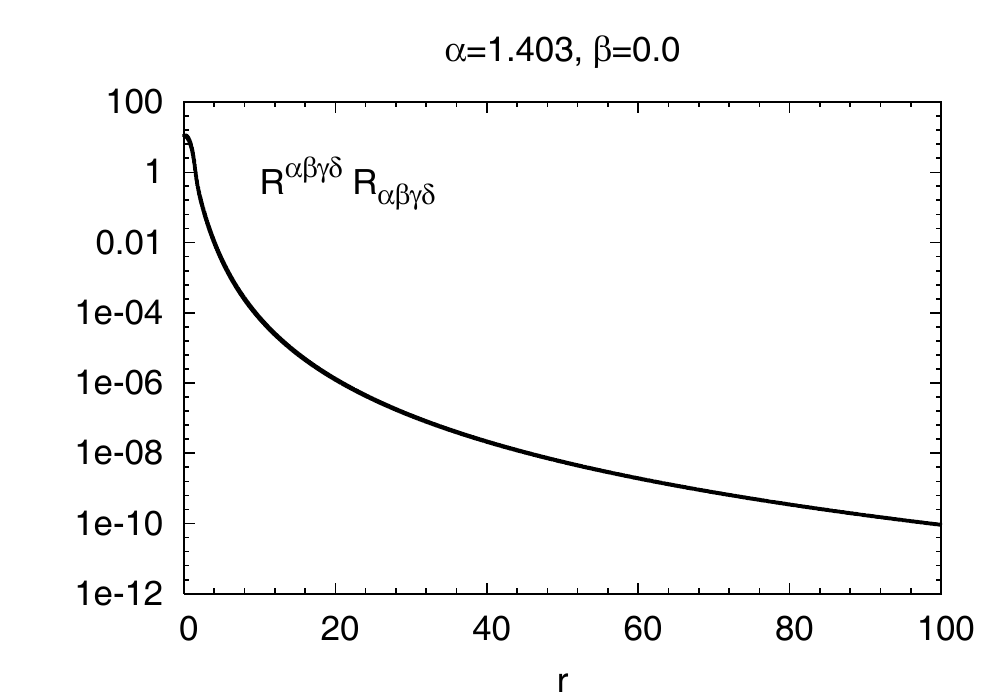}}
\end{center}
\vspace{-0.8cm}
\caption{The metric functions $-g_{tt}$ (a) and $1/g_{rr}$ (b),
the potential $U$ (c) and the curvature invariant $K$ (d)
versus the radial coordinate $r$
for $(\alpha_{\rm max},\beta)=(1.403,0)$. \label{series2a}}
\end{figure}

We now consider the orbits for the maximal value of $\alpha$,
i.e., for the parameters $(\alpha_{\rm max},\beta)=(1.403,0)$.
Again, the metric coefficient $-g_{tt}$ is a
monotonically rising function,
and the metric coefficient $1/g_{rr}$
exhibits a minimum at some
value of the radial coordinate, $r_{0}$,
although the minimum is much deeper now.
The Kretschmann scalar is still finite everywhere and not too large.

\begin{figure}[t]
\begin{center}
\subfigure[][inner region low energy]{\includegraphics[width=0.3\textwidth]{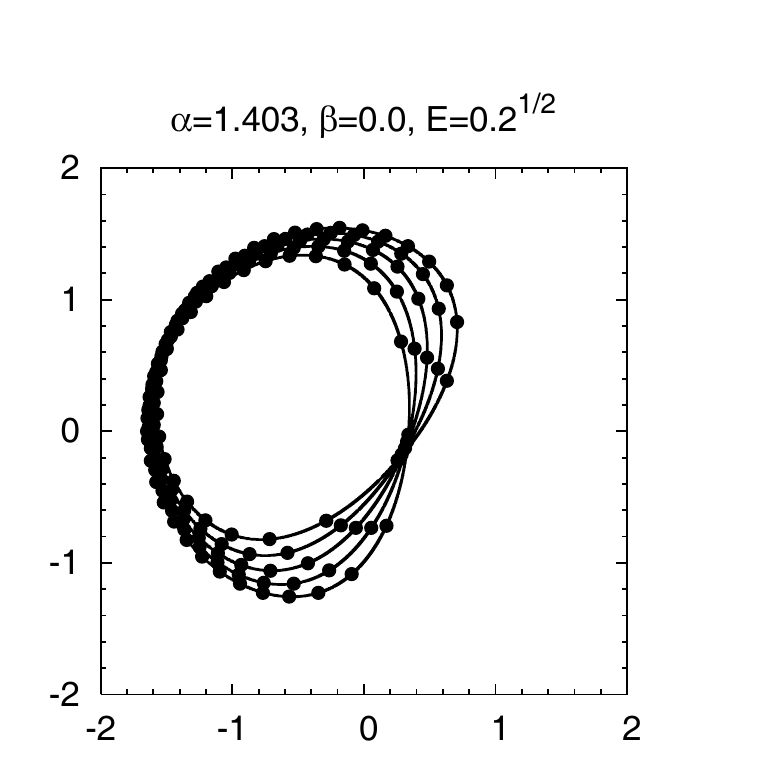}}
\subfigure[][inner region higher energy]{\includegraphics[width=0.3\textwidth]{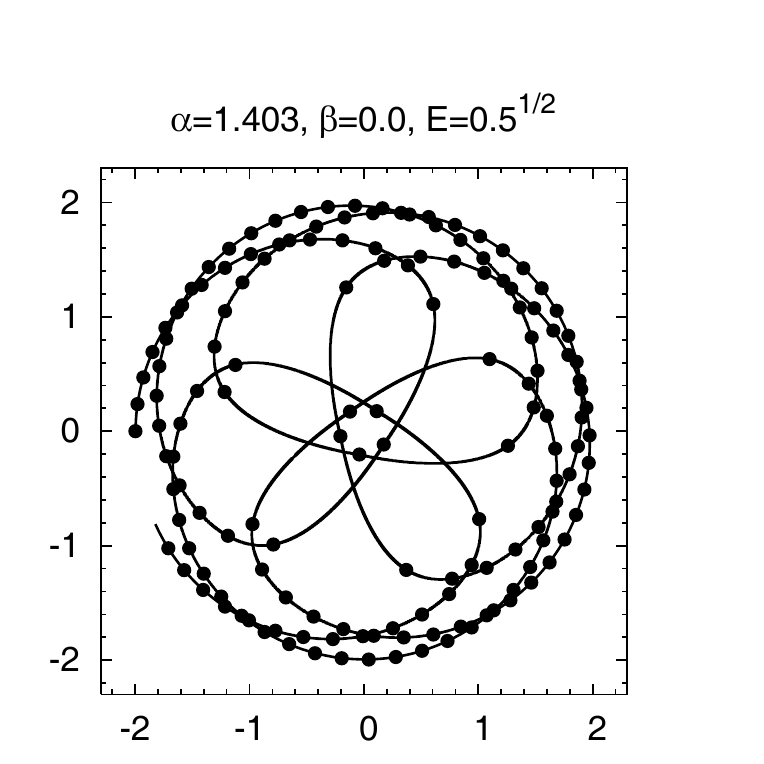}}
\subfigure[][inner region $E = \sqrt{0.908}$]{\includegraphics[width=0.3\textwidth]{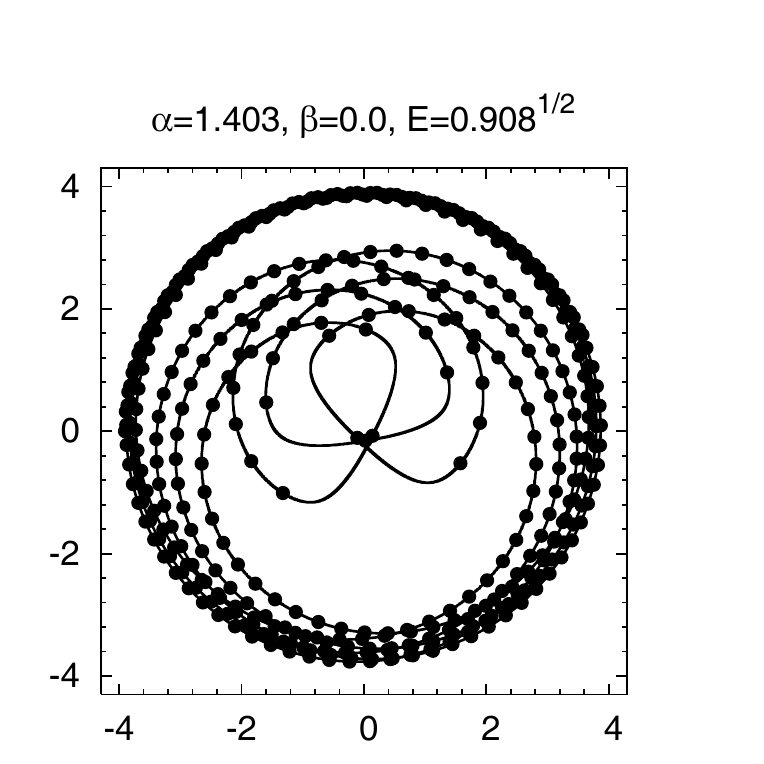}}

\subfigure[][outer region $E = \sqrt{0.908}$]{\includegraphics[width=0.3\textwidth]{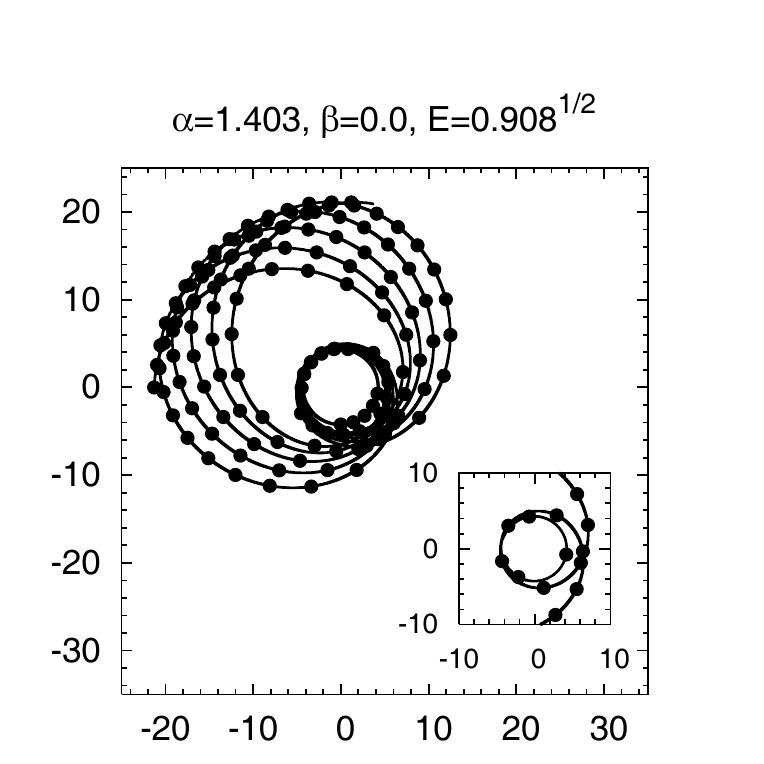}}
\subfigure[][highly elliptic orbit in outer region]{\includegraphics[width=0.3\textwidth]{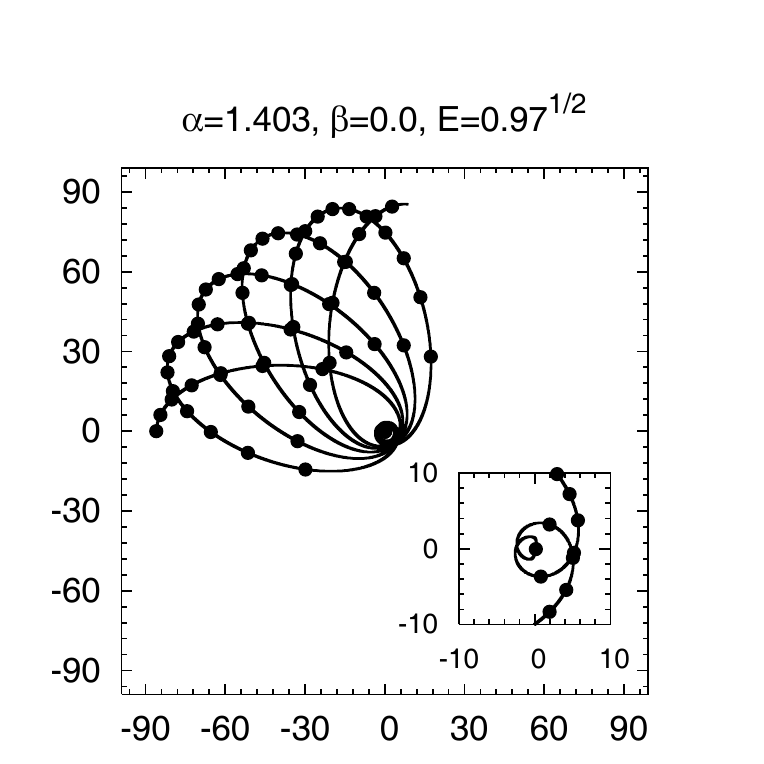}}
\subfigure[][escape orbit]{\includegraphics[width=0.3\textwidth]{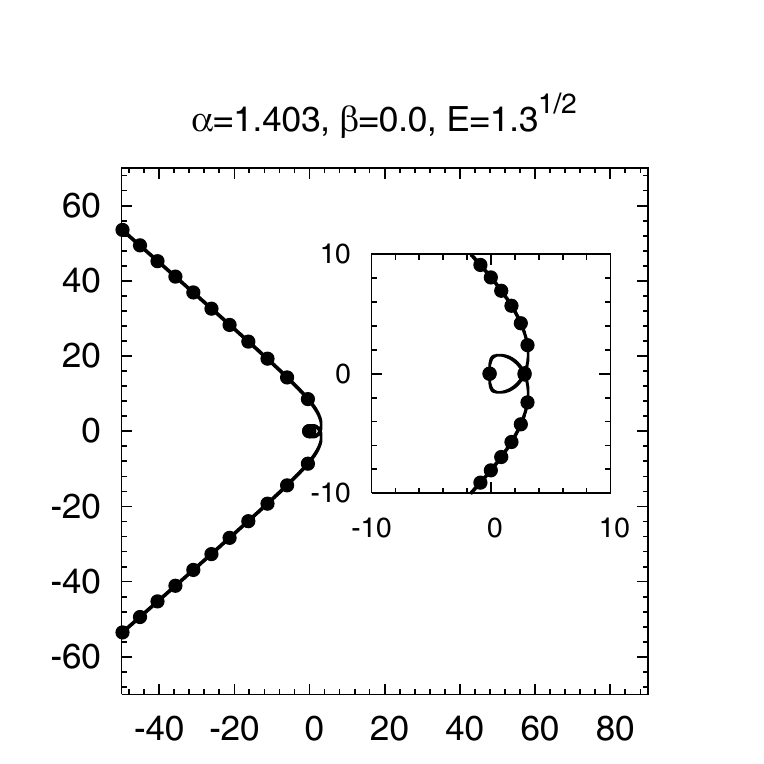}}
\end{center}
\caption{
Particle orbits $r(\varphi)$ for $(\alpha_{\rm max},\beta)=(1.403,0)$.
The crosses indicate units of elapsed proper time.
\label{series2b}}
\end{figure}

The most interesting feature here is that the potential $U$
can now possess two minima and a maximum.
In such a case, bound orbits are present in two distinct regions of space.
We exhibit typical examples of such bound orbits
in Fig.~\ref{series2b} for an angular momentum of $L=4.3$.
The first three figures (a), (b) and (c) show orbits in the inner region
(associated with the inner minimum),
for a low energy value, $E=\sqrt{0.2}$,
for a higher energy value, $E=\sqrt{0.5}$,
and for the (all but) highest value possible for
a bound orbit in the inner region, $E=\sqrt{0.908}$.

For very low energies the orbits of course approach a circular orbit.
As the energy increases, the orbits have a deformed shape,
and show a very pronounced perihelion shift ((a) and (b)).
When the limiting energy value $E_{\rm max}$
for bound orbital motion in the inner region is approached,
the associated unstable circular orbit becomes apparent,
since it forms the outer envelope of the trajectory (c).

In the energy interval between the outer minimum and the maximum,
there are orbits in the inner region (c) as well as in the outer region (d).
Here the unstable circular orbit associated with the maximum
becomes apparent as the inner boundary for the orbital motion.
For still higher energies, in the range $E_{\rm max} < E < 1$
one observes quasi--elliptic orbits (e).
In the outer region these look like ellipses.
When they get close to the inner region,
the unstable circular orbit is again reflected in the motion,
but it no longer forms a boundary.
Instead the motion proceeds further inwards,
orbits close to the center and then proceeds outwards again,
as highlighted in the inset of (e).

For energies $E > 1$, we observe scattering states (f).
The particle then encircles the central region several times
(twice in (f))
before escaping into infinity again.


\begin{figure}[th!]
\begin{center}
\subfigure[][$-g_{tt}$]{\includegraphics[width=0.4\textwidth]{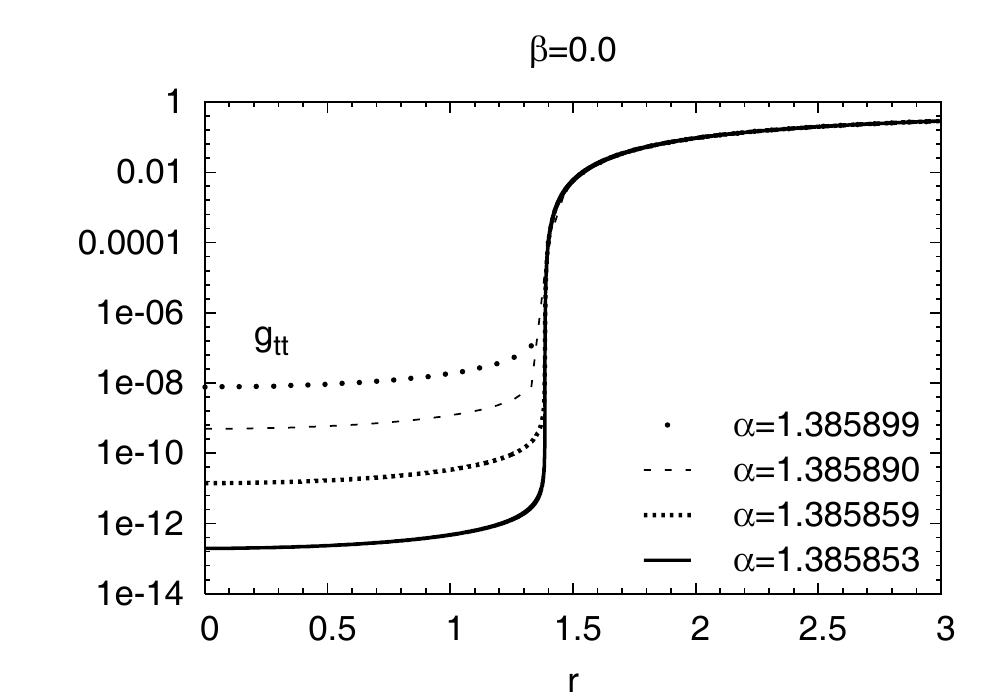}}
\subfigure[][$1/g_{rr}$]{\includegraphics[width=0.4\textwidth]{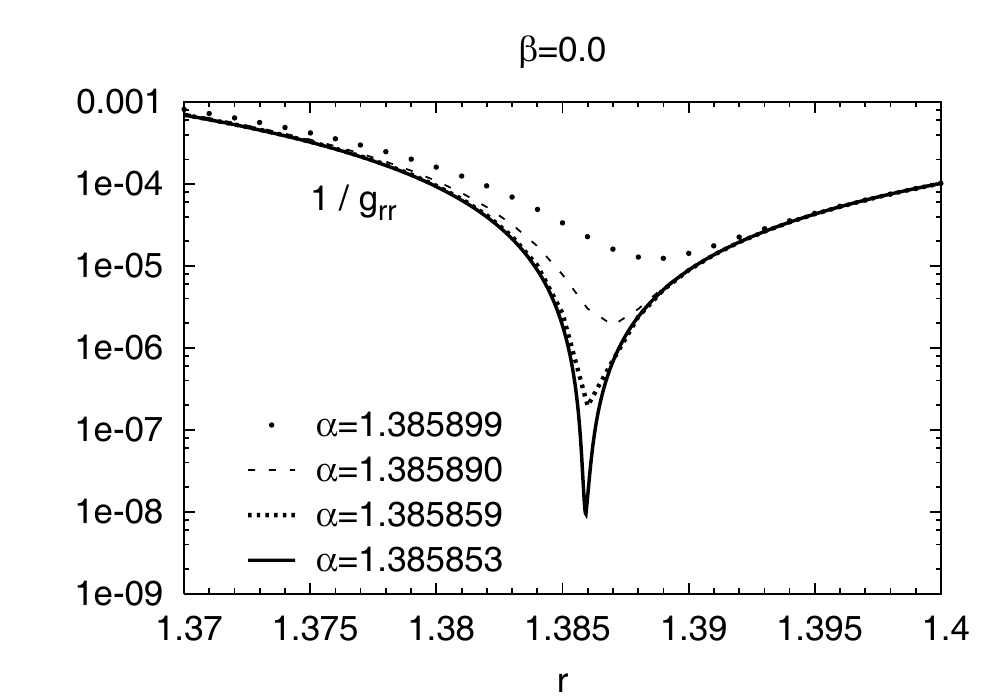}}
\subfigure[][$K$]{\includegraphics[width=0.4\textwidth]{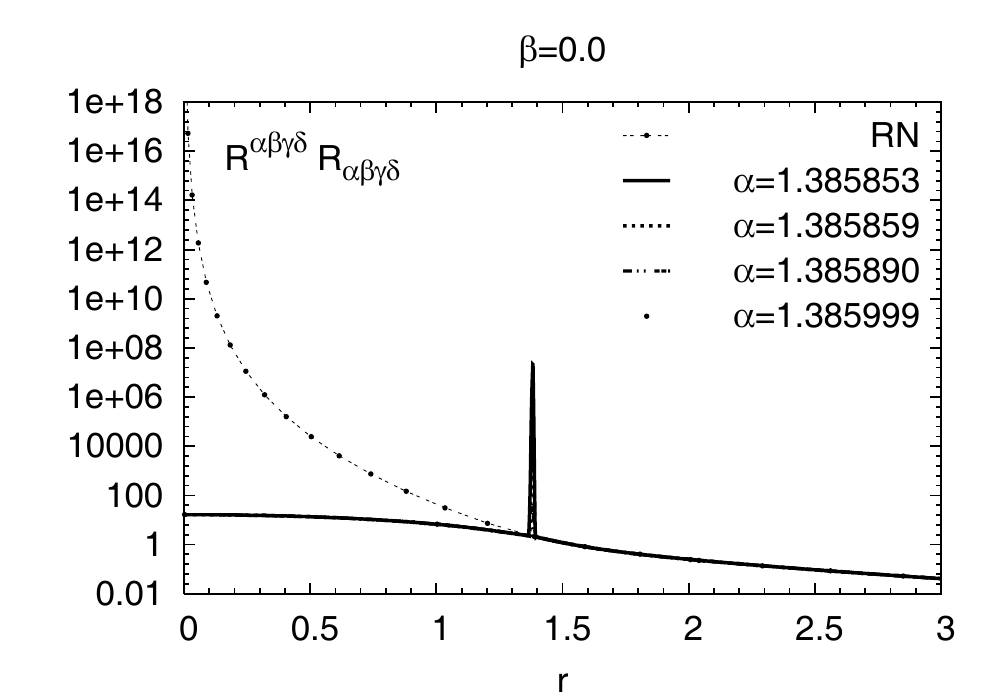}}
\subfigure[][$K$]{\includegraphics[width=0.4\textwidth]{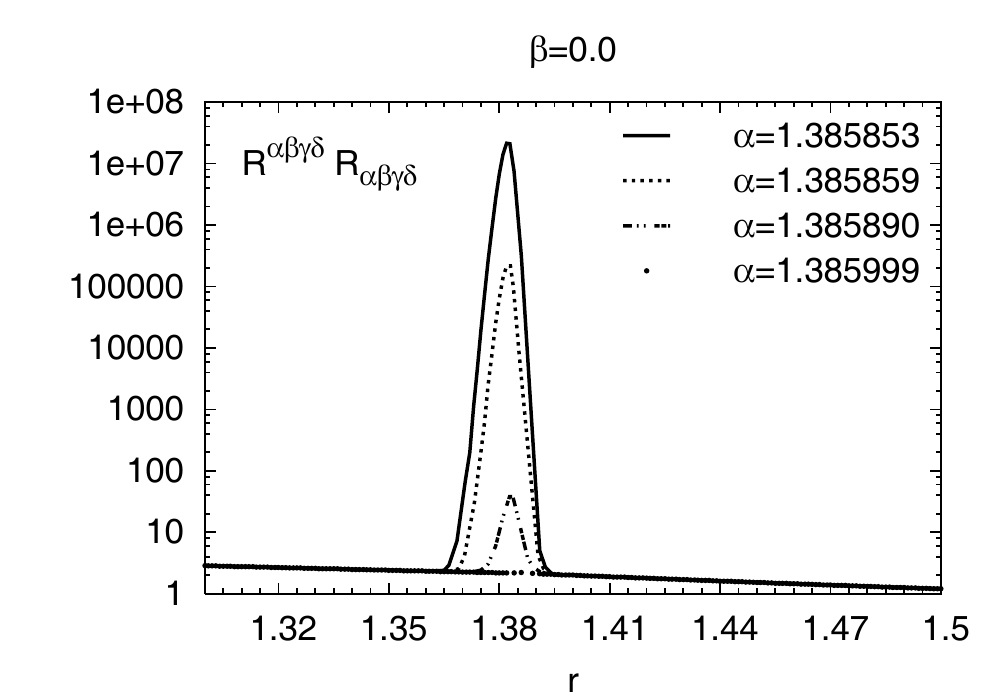}}
\subfigure[][$U$]{\includegraphics[width=0.4\textwidth]{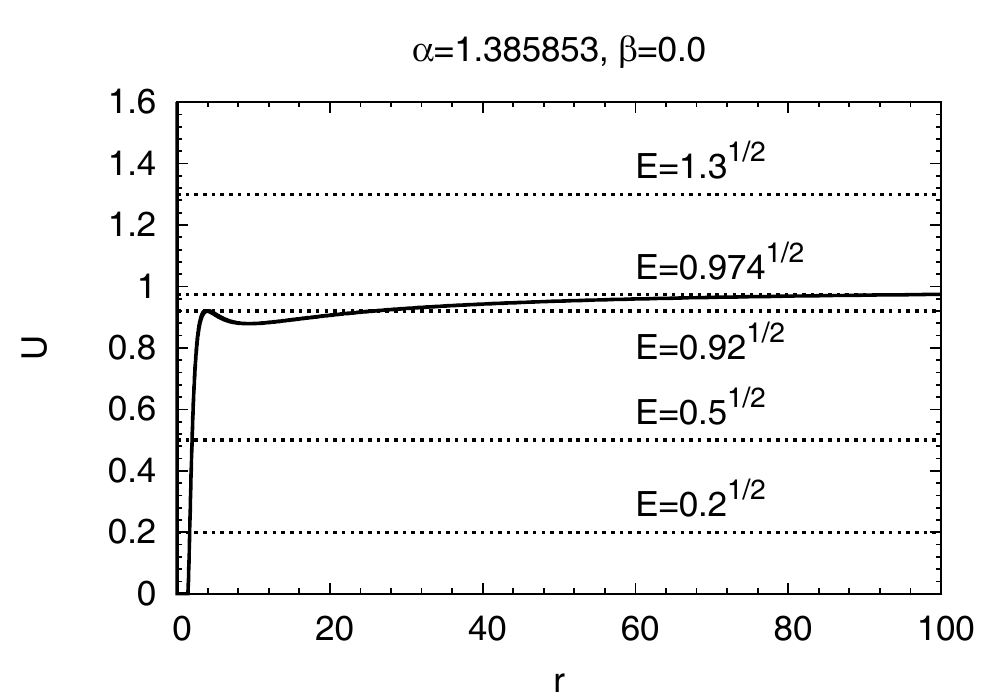}}
\subfigure[][$U_{\rm eff}$]{\includegraphics[width=0.4\textwidth]{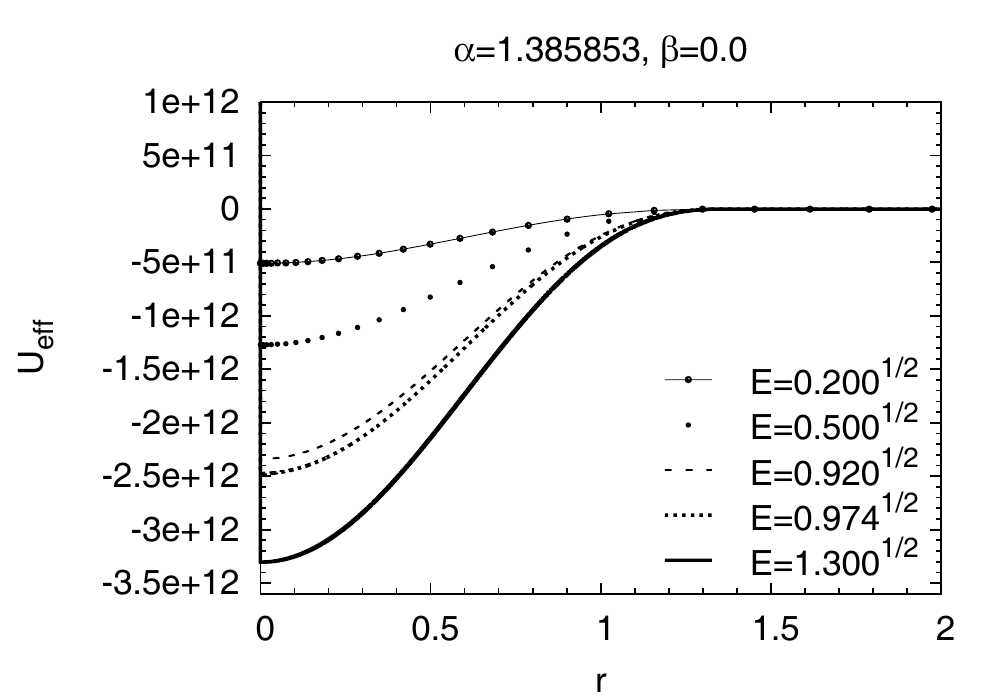}}
\end{center}
\caption{The metric functions $-g_{tt}$ (a) and $1/g_{rr}$ (b),
the curvature invariant $K$ (c) and (d),
the potential $U$ (e), and the effective potential $U_{\rm eff}$ (f),
versus the radial coordinate $r$
for $(\alpha_{\rm cr},\beta)=(1.38585,0)$.
(RN denotes the extremal Reissner--Nordstr\"om black hole.)
\label{series3a}}
\end{figure}

\boldmath
\subsubsection{Orbits at $\alpha_{\rm cr} = 1.38585$}
\unboldmath

We now demonstrate, how the monopole space--time
evolves as the critical solution is approached.
As $\alpha \rightarrow \alpha_{\rm cr}$
the minimum of the metric function $1/g_{rr}$
decreases and reaches zero in the limit.
This decrease towards zero is seen in Fig.~\ref{series3a}.
Note, that $\alpha_{\rm cr}$ is here determined only
within a certain numerical accuracy,
and therefore the final solution obtained is only {\sl almost} critical:
$1/g_{rr}$ does not yet fully reach zero at some coordinate value
$r_0$, 
but it is extremely small there.

The true limiting solution then consists of two parts:
the interior region $0 \le r < r_0$
and the exterior region $r_0 < r < \infty$.
Since the limiting exterior solution
corresponds to an extremal Reissner--Nordstr\"om space--time,
$r_0= \alpha$ must hold for the limiting solution,
according to Eq.~(\ref{RNex}).
The limiting interior solution in a non-Abelian solution
and regular at the origin.

As $\alpha \rightarrow \alpha_{\rm cr}$
the metric function $-g_{tt}$ becomes increasingly small
in the inner region, $0 \le r < r_0$,
tending to zero in the limit.
Close to $r_0$, however, $-g_{tt}$ rises very steeply
to assume its asymptotic value of one,
as imposed by the boundary conditions.
In the limit $\alpha \rightarrow \alpha_{\rm cr}$,
$-g_{tt}$ appears to become singular at $r_0$.

The emergence of a singularity at $r_0$
is also indicated by the Kretschmann scalar of the monopole space--time,
which appears to diverge in the limit.
In contrast, the Kretschmann scalar of the associated
extremal Reissner--Nordstr\"om black hole remains finite
at its degenerate horizon $r_0$, but diverges for
$r \rightarrow 0$, as seen in Fig.~\ref{series3a}.

\begin{figure}[t]
\begin{center}
\subfigure[][inner region, $E=0.2^{1/2}$]{\includegraphics[width=0.3\textwidth]{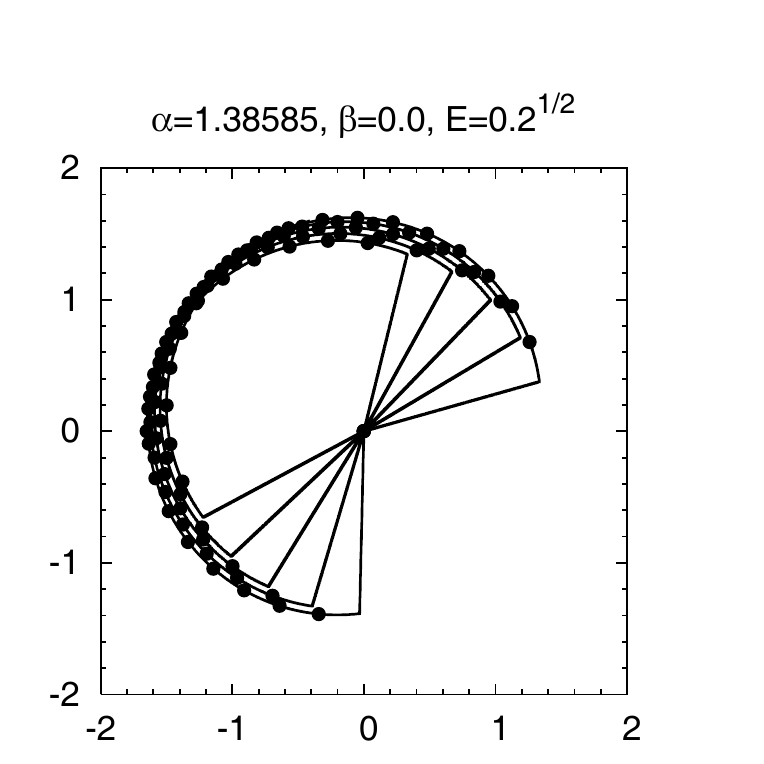}}
\subfigure[][inner region, $E=0.5^{1/2}$]{\includegraphics[width=0.3\textwidth]{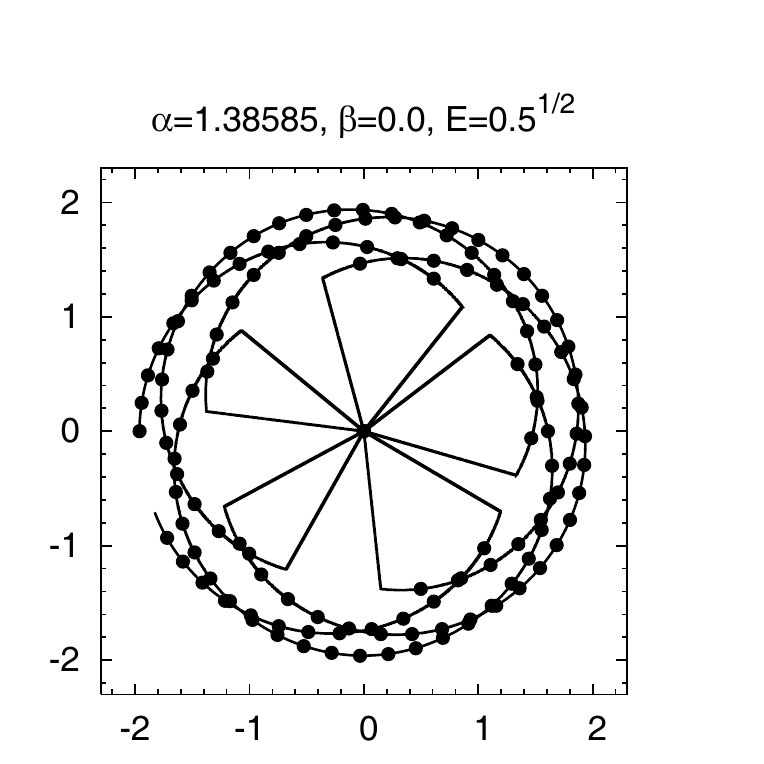}}
\subfigure[][inner region, $E=0.92^{1/2}$]{\includegraphics[width=0.3\textwidth]{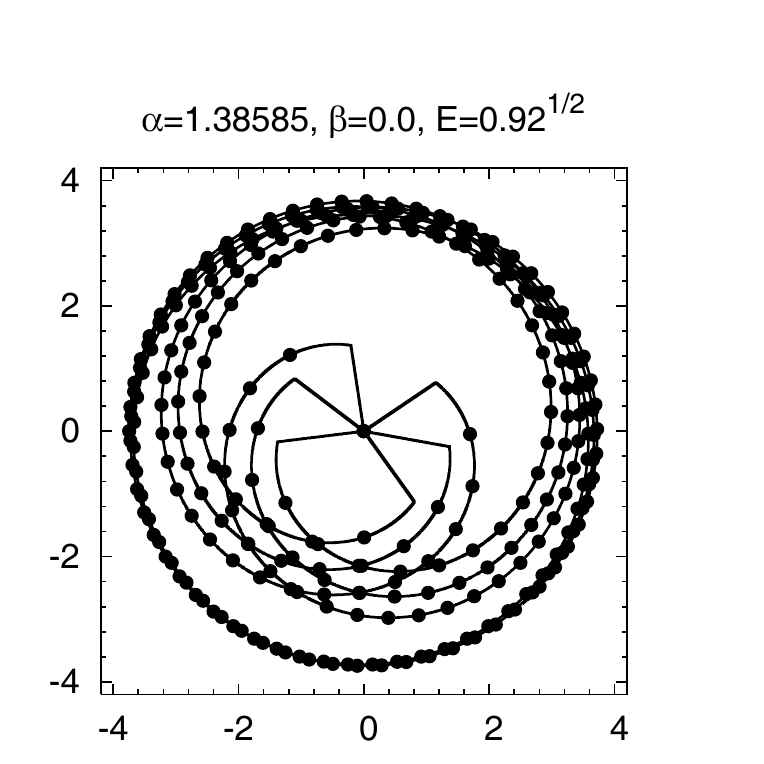}}

\subfigure[][outer region, $E=0.92^{1/2}$]{\includegraphics[width=0.3\textwidth]{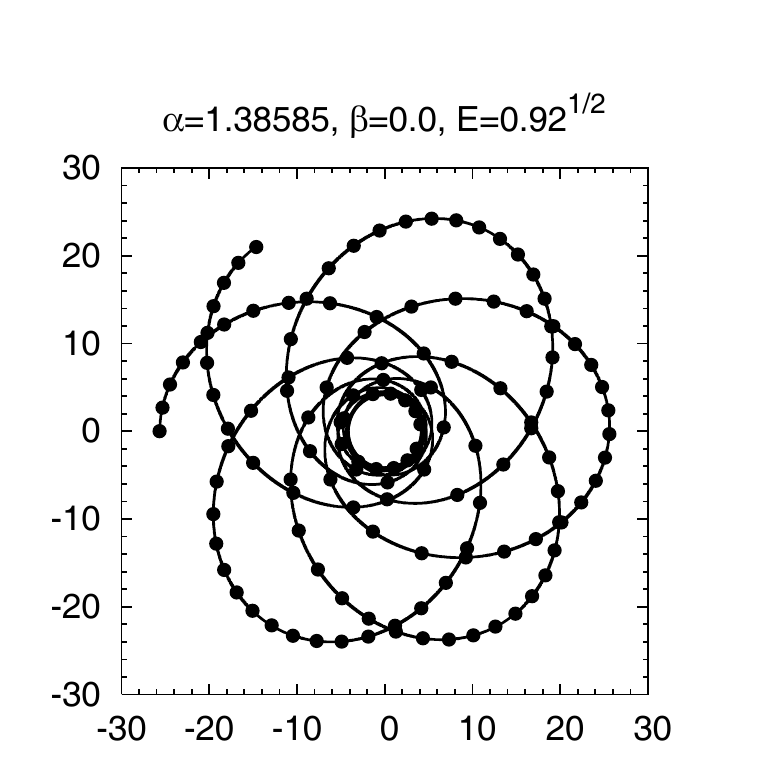}}
\subfigure[][outer region, $E=0.974^{1/2}$]{\includegraphics[width=0.3\textwidth]{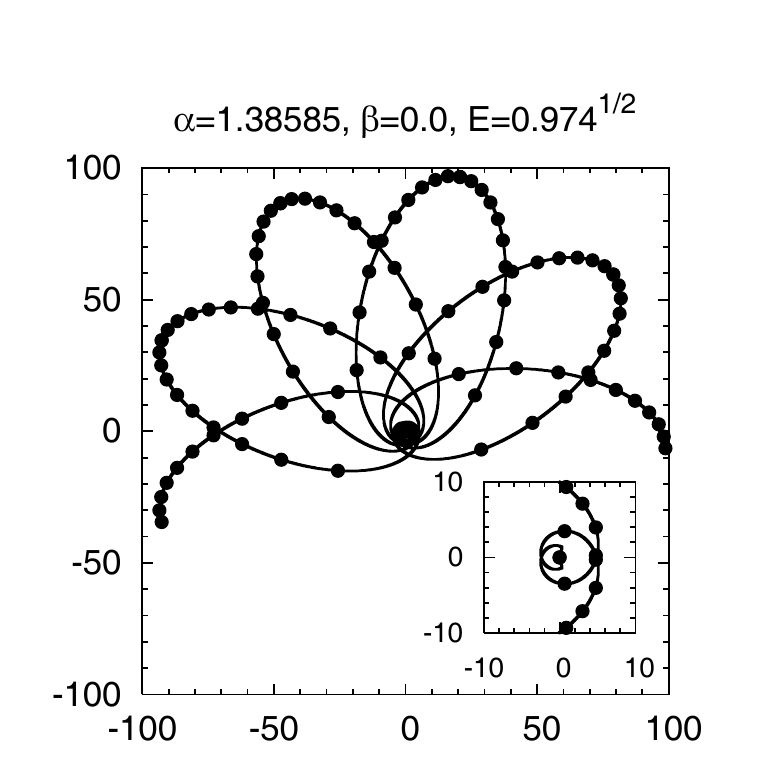}}
\subfigure[][escape orbit, $E = 1.3^{1/2}$]{\includegraphics[width=0.3\textwidth]{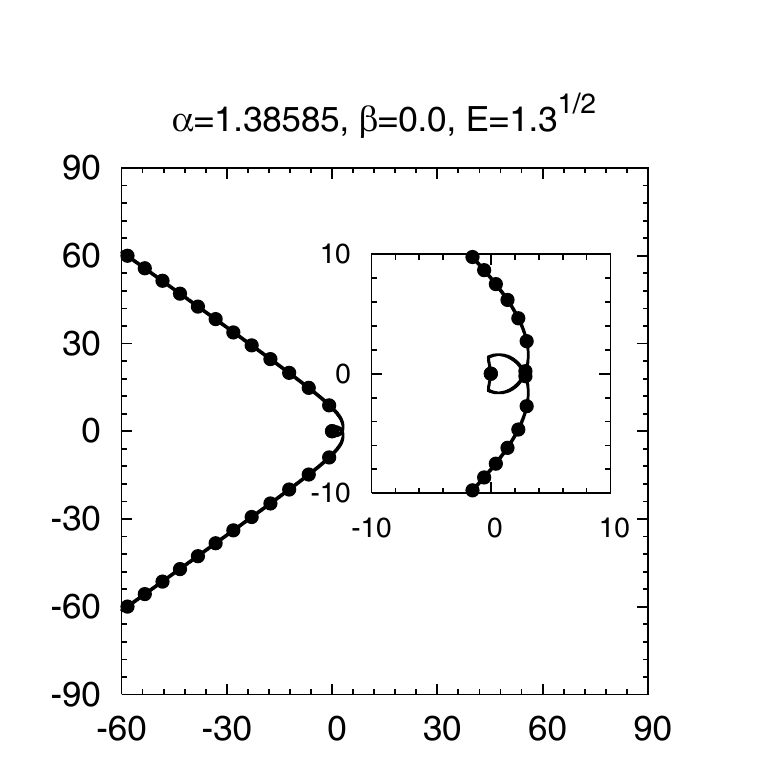}}
\end{center}
\caption{
Particle orbits $r(\varphi)$ for $(\alpha_{\rm cr},\beta)=(1.385853,0)$.
The dots indicate units of elapsed proper time.
\label{series3b}}
\end{figure}

Let us now consider the motion of particles in the
{\sl almost} critical space--time,
with parameters $(\alpha,\beta) = (1.38585,0)$,
and exemplify the intriguing effects of this {\sl almost} critical space--time
on the orbital motion.
We study the orbits
for a representative value of the angular momentum, $L=4.3$.
As shown in Fig.\ref{series3a},
the potential $U$ then again possesses two minima.
We therefore expect the same general types of orbits
as in the previous case.

We display a set of characteristic orbits in Fig.~\ref{series3b}.
The bound orbits in the inner region (a) and (b)
(connected to the inner minimum of the potential)
are now hugely distorted as compared to ordinary bound orbits.
In the energy interval between the outer minimum and the maximum,
bound orbits are present in the inner (c) as well as in the outer region (d).
Again the unstable circular orbit associated with the maximum
becomes apparent as the outer envelope of the allowed motion
in the inner region (c)
and as the inner boundary for the trajectories in the outer region (d).
For still higher energies, in the range $E_{\rm max} < E < 1$
we again observe quasi--elliptic orbits with additional inner
loops circling the center (e),
while for energies $E > 1$, we again observe scattering states (f).

The most surprising feature
is the deformation of the orbits in the inner region.
As seen in Fig.~\ref{series3a},
the orbits evolve smoothly until they reach the vicinity
of the critical value of the radial coordinate, $r_0$.
There they abruptly change direction
and approach in almost straight radial lines the center.
Very close to the center they are reflected
(passing partial arcs),
and then move outwards again in almost straight radial lines,
until they reach the vicinity of $r_0$,
where they again abruptly change direction.

This intriguing pattern of movement can be understood as follows:
for $r < r_0$ the effective potential very quickly tends to
extremely large negative values, as seen in Fig.~\ref{series3a}.
Therefore in this region a particle is vigorously
attracted towards the origin.
This huge attraction leads to orbits which are almost straight radial lines.
(Note, that this interpretation is related to the first column
of the table in subsection \ref{subsec:independentgttgrr}.)
The deflection close to the center is caused by
the potential barrier there.

Interesting is also a glance at the proper time of a particle
along these almost straight radial parts of the orbits.
Since $-g_{tt}$ is extremely small, a particle needs almost
no proper time to traverse the inner region,
as exemplified in Fig.~\ref{series3a}
(and also discussed in subsection \ref{subsec:independentgttgrr}).

\begin{figure}[t]
\begin{center}
\subfigure[][$l(r)$ for various $\alpha$]{\includegraphics[width=0.4\textwidth]{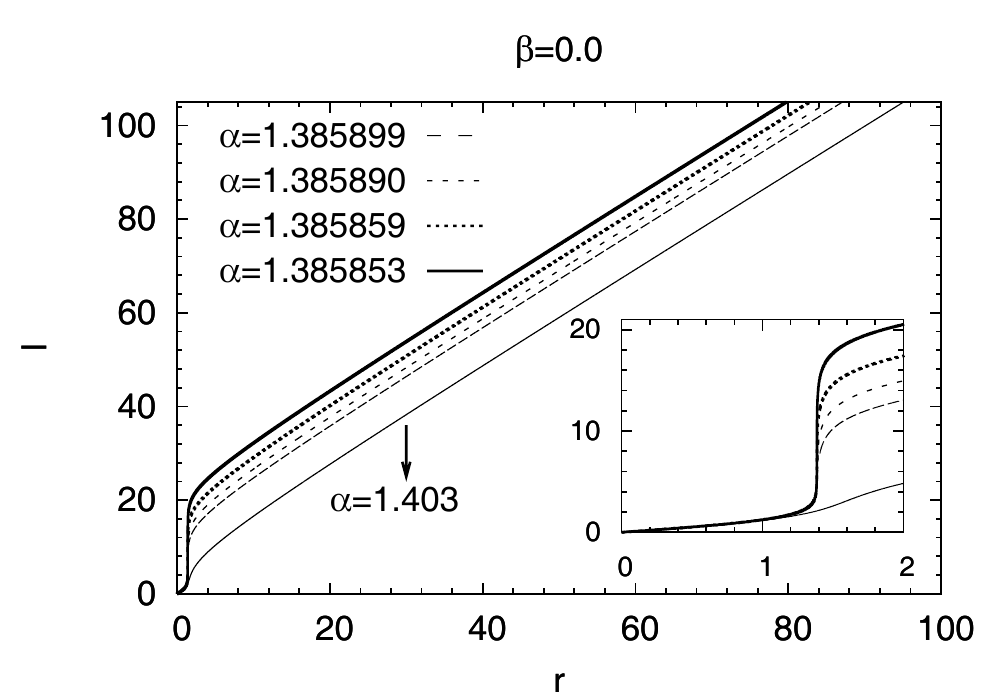}}
\subfigure[][$l(r)$ for $\alpha$ almost $\alpha_{\rm crit}$]{\includegraphics[width=0.4\textwidth]{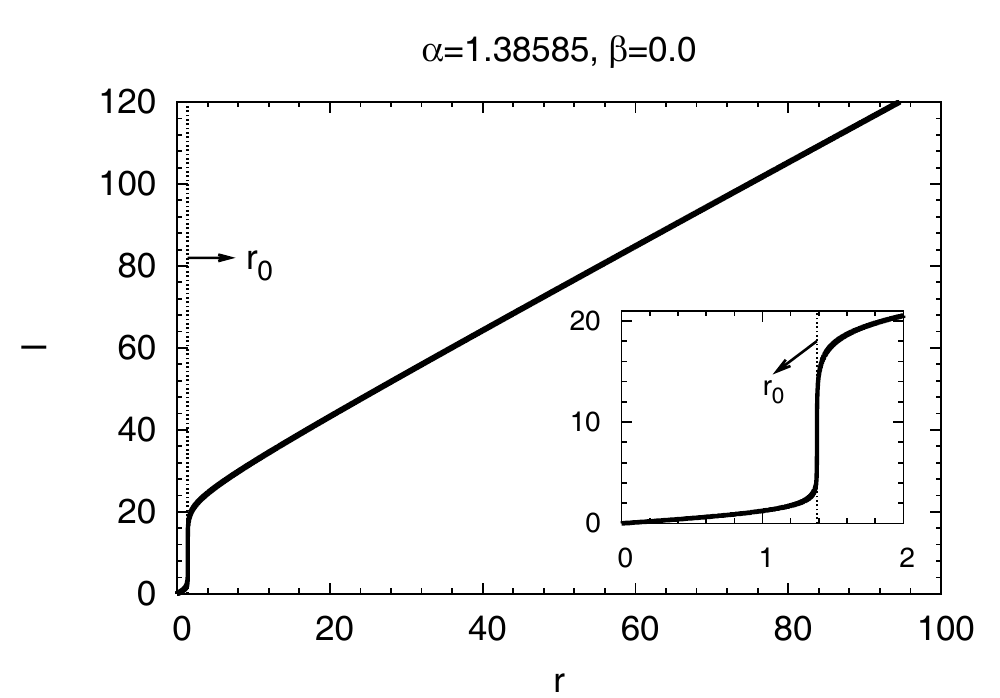}}
\end{center}
\caption{The measured spatial distance $l$ versus the radial coordinate $r$. As $\alpha$ approaches $\alpha_{\rm crit}$ a throat develops at $r = r_0$. \label{series6a}}
\end{figure}

Let us finally consider these orbits in terms of the proper distance
$l$ instead of the radial coordinate $r$.
In Fig.~\ref{series6a} we exhibit the dependence $l(r)$
for a sequence of solutions approaching the critical solution.
The figure demonstrates, that the space--time develops a throat,
as $\alpha \rightarrow \alpha_{\rm cr}$.
In the limit, the throat becomes infinitely long \cite{gmono}.
For the {\sl almost} critical value
$\alpha=1.385853$, $l$ grows already by a factor of 20 in the vicinity
of $r_0$.

\begin{figure}[t]
\subfigure[][]{\includegraphics[width=0.3\textwidth]{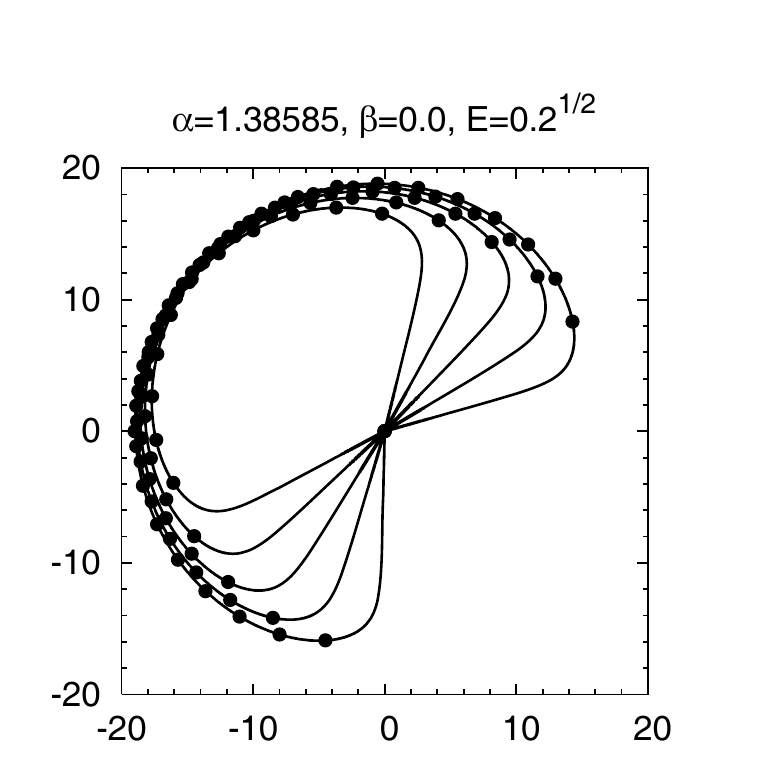}}
\subfigure[][]{\includegraphics[width=0.3\textwidth]{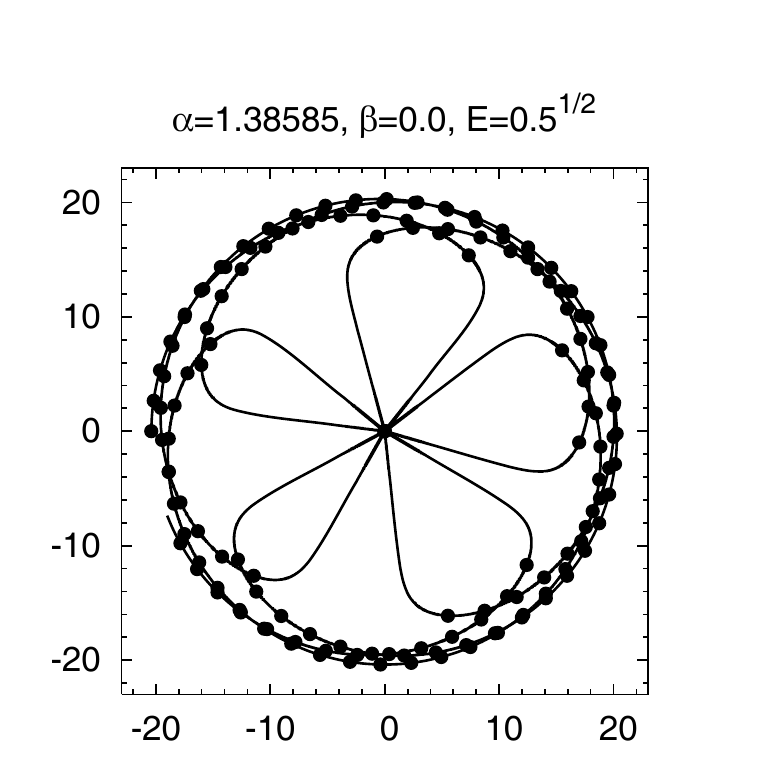}}
\subfigure[][]{\includegraphics[width=0.3\textwidth]{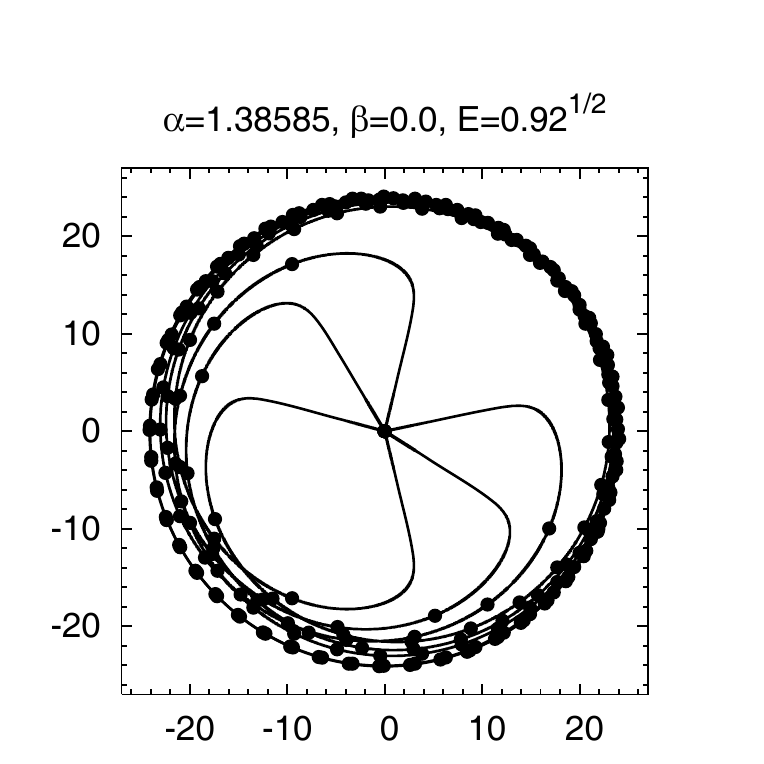}}
\caption{Particle orbits $l(\varphi)$ for $(\alpha_{\rm cr},\beta)=(1.385853,0)$.
The dots indicate units of elapsed proper time. \label{series6b}}
\end{figure}

The rapid growth of $l$ in the vicinity of $r_0$ has
significant influence on the shape of the orbits.
The abrupt directional changes seen in the orbits $r(\varphi)$
in the vicinity of $r_0$
are smoothed out, when $l(\varphi)$ is considered instead,
since now the orbits are seen to move within the long throat
of the space-time.
This is demonstrated in Fig.~\ref{series6b}
for the inner orbits of Fig.~\ref{series3a}.

\boldmath
\subsection{Orbits for $\beta>0$}
\unboldmath

\begin{figure}[t]
\begin{center}
\subfigure[][$-g_{tt}$]{\includegraphics[width=0.4\textwidth]{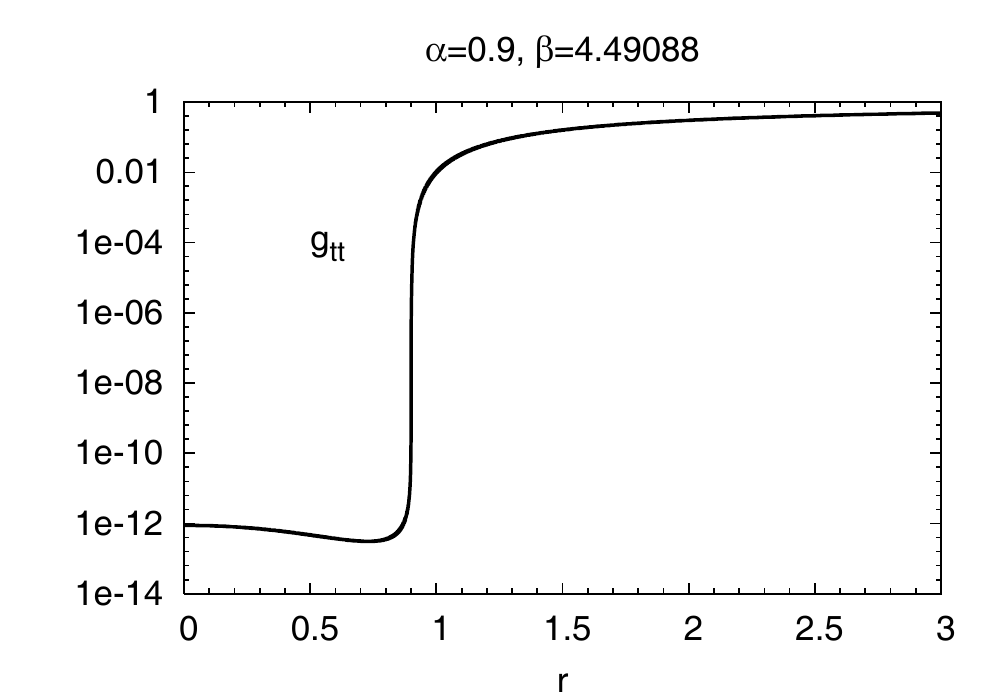}}
\subfigure[][$1/g_{rr}$]{\includegraphics[width=0.4\textwidth]{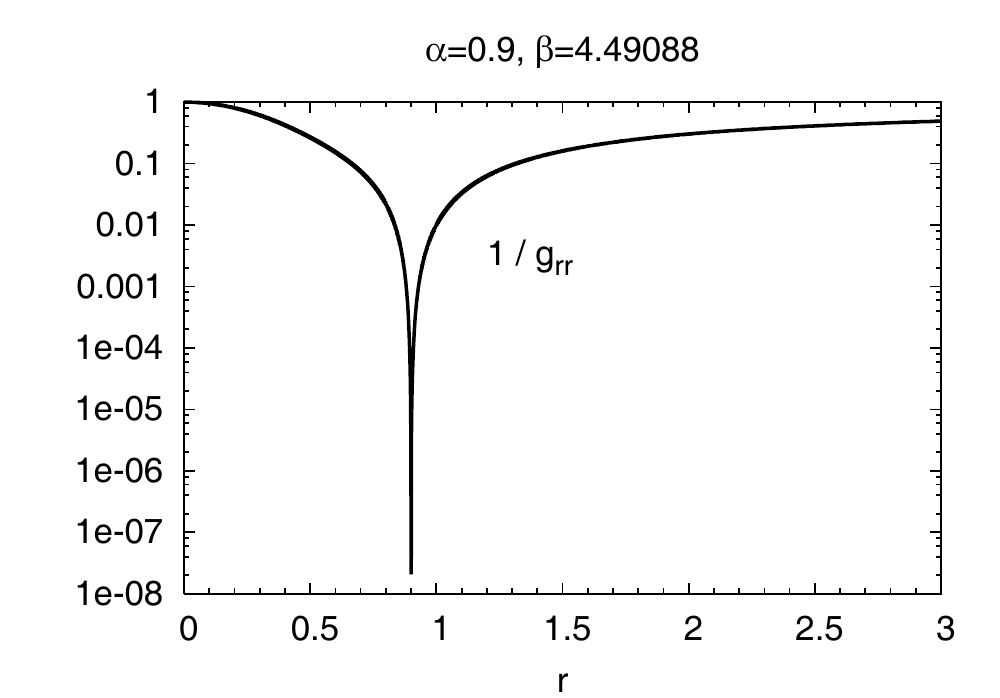}}
\subfigure[][$U$]{\includegraphics[width=0.4\textwidth]{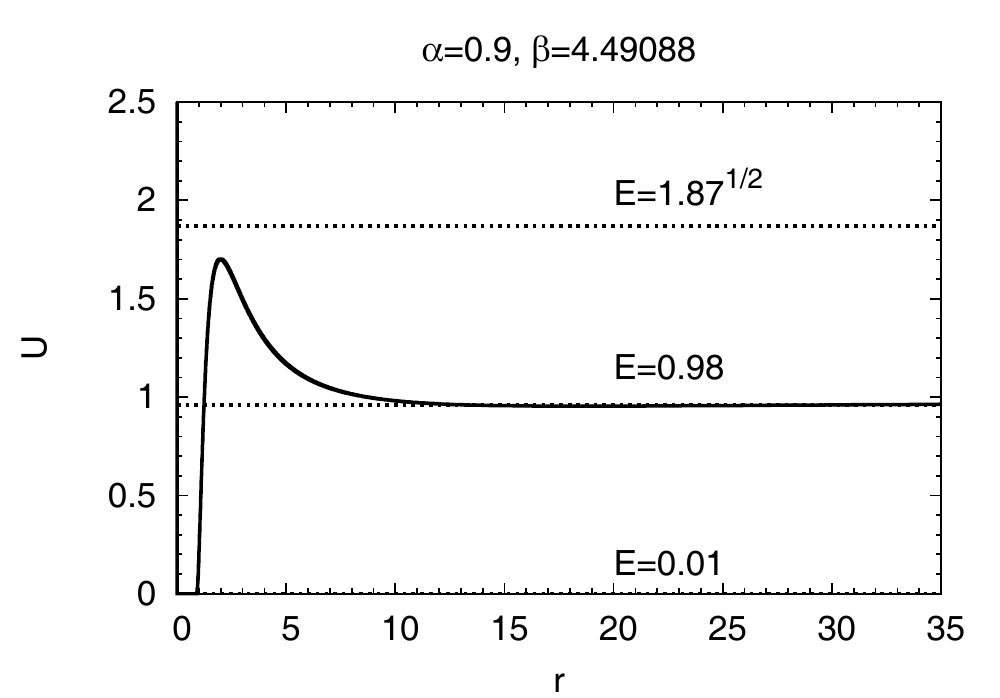}}
\subfigure[][$K$]{\includegraphics[width=0.4\textwidth]{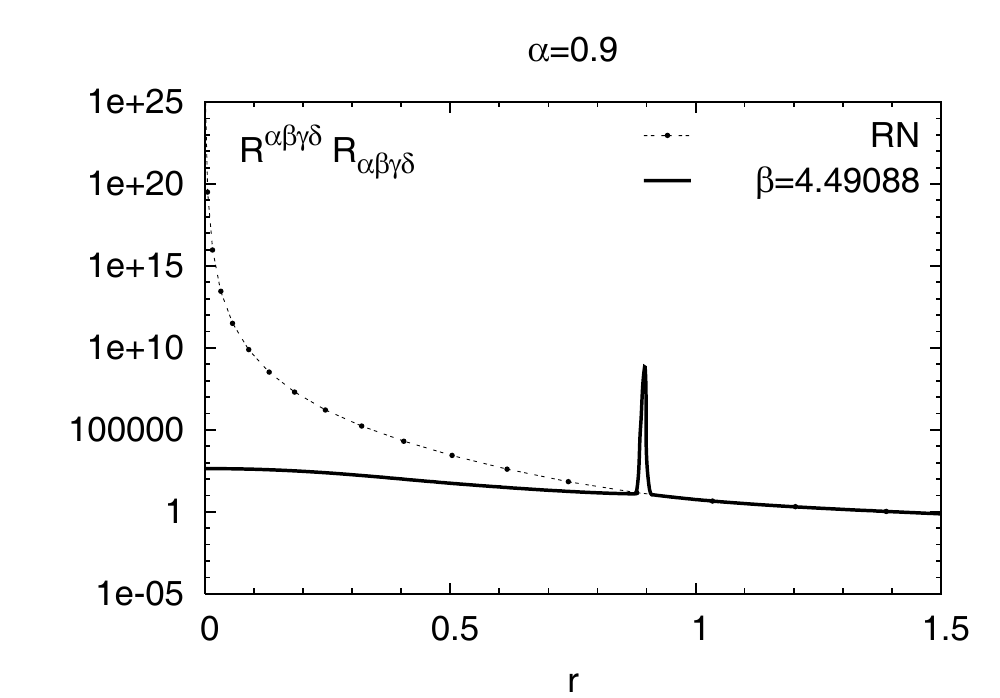}}
\end{center}
\caption{The metric functions $-g_{tt}$ (a) and $1/g_{rr}$ (b),
the potential $U$ (c),
and the curvature invariant $K$ (d)
versus the radial coordinate $r$
for $(\alpha,\beta)=(0.9,4.49088)$.
(RN denotes the extremal Reissner--Nordstr\"om black hole.)
\label{series4a}}
\end{figure}


Let us now consider the effect of a finite value of
the parameter $\beta$ and thus the Higgs mass
on the monopole space--times and
consequently on the orbits in these space--times.

As $\beta$ increases $\alpha_{\rm max}$ decreases.
At the same time, $\alpha_{\rm cr}$ and $\alpha_{\rm max}$
approach each other, until they merge.
The critical space--time then arises at the maximal
possible value of $\alpha$ for the given value of $\beta$.
We exhibit such an {\sl almost} critical space--time
at $(\alpha,\beta)=(0.9,4.49088)$
and study its orbits.

As $\beta$ increases further another interesting phenomenon
arises: $1/g_{rr}$ develops two minima.
We study the evolution of such space--times towards
the corresponding critical space--time
at $(\alpha,\beta)=(0.85,8.282558)$.
We discuss the effects on the orbits of particles,
and we exhibit orbits of light rays.

\boldmath
\subsubsection{Orbits at $\alpha=0.9$, $\beta=4.49088$}
\unboldmath

We now address motion in an {\sl almost} critical
space--time at a finite, but still small value of $\beta$,
namely at $\beta=4.49088$.
This value of $\beta$ is obtained, by fixing $\alpha=0.9$
and then increasing $\beta$, until $\alpha=0.9$ becomes the maximal
and at the same time critical value, $\alpha_{\rm max}=\alpha_{\rm cr}=0.9$,
i.e., for this specific value $\beta=4.49088$ no regular solutions
exist beyond $\alpha=0.9$.

\begin{figure}[t]
\begin{center}
\subfigure[][inner region, $E=10^{-5}$]{\includegraphics[width=0.3\textwidth]{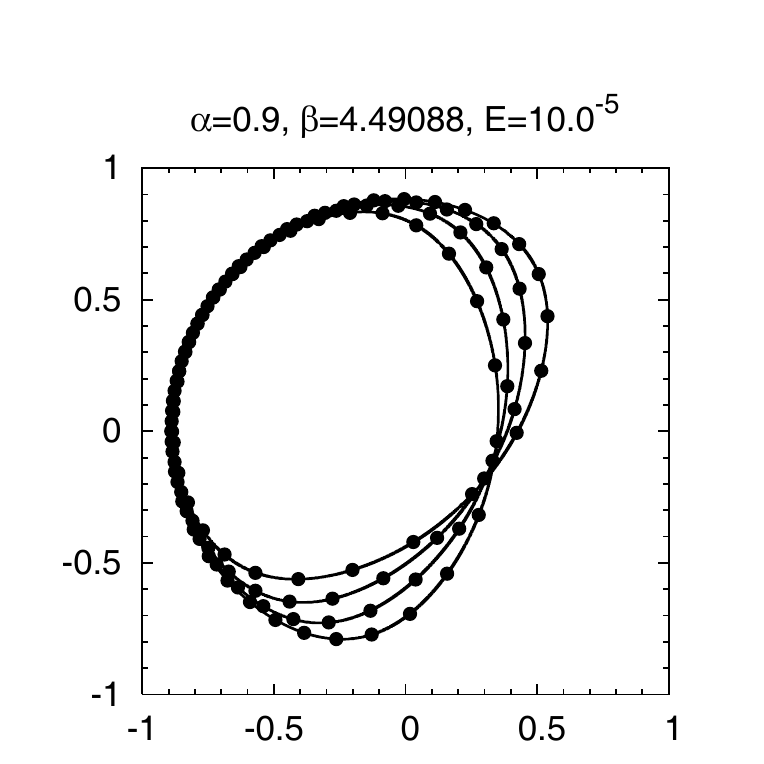}}
\subfigure[][inner region, $E=10^{-2}$]{\includegraphics[width=0.3\textwidth]{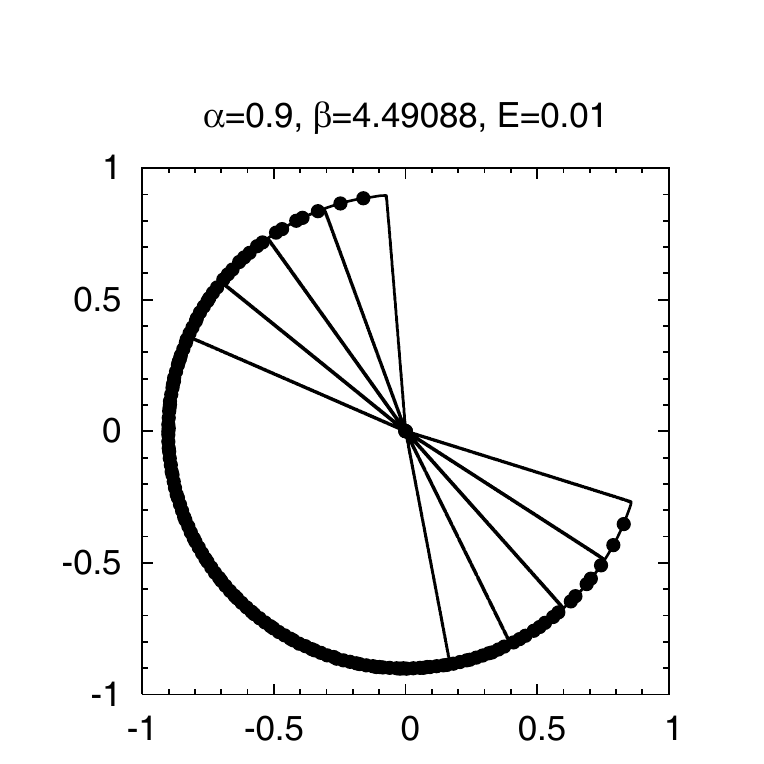}}
\subfigure[][inner region, $E=0.98$, inner region]{\includegraphics[width=0.3\textwidth]{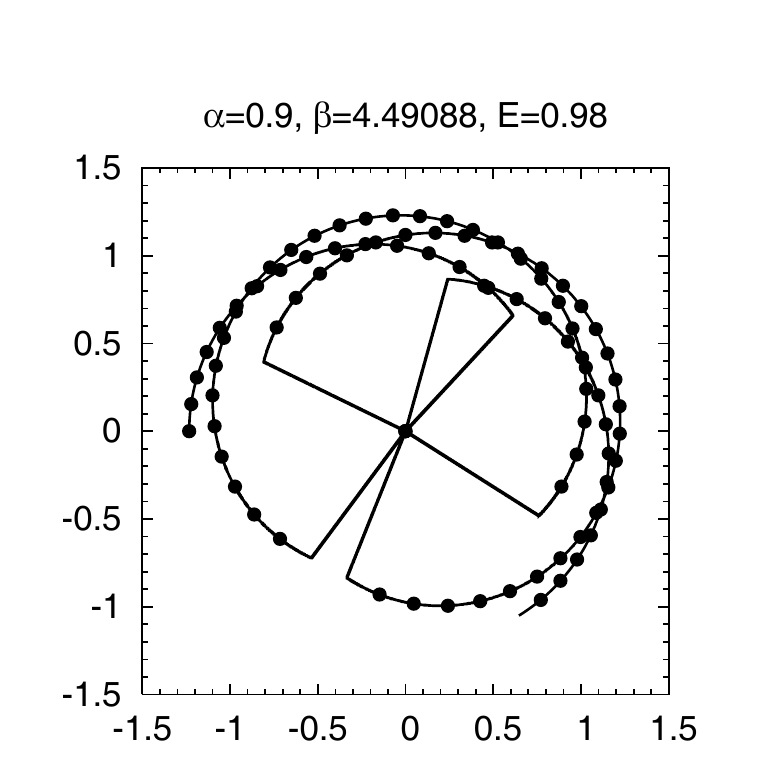}}

\subfigure[][outer region, $E=0.98$]{\includegraphics[width=0.3\textwidth]{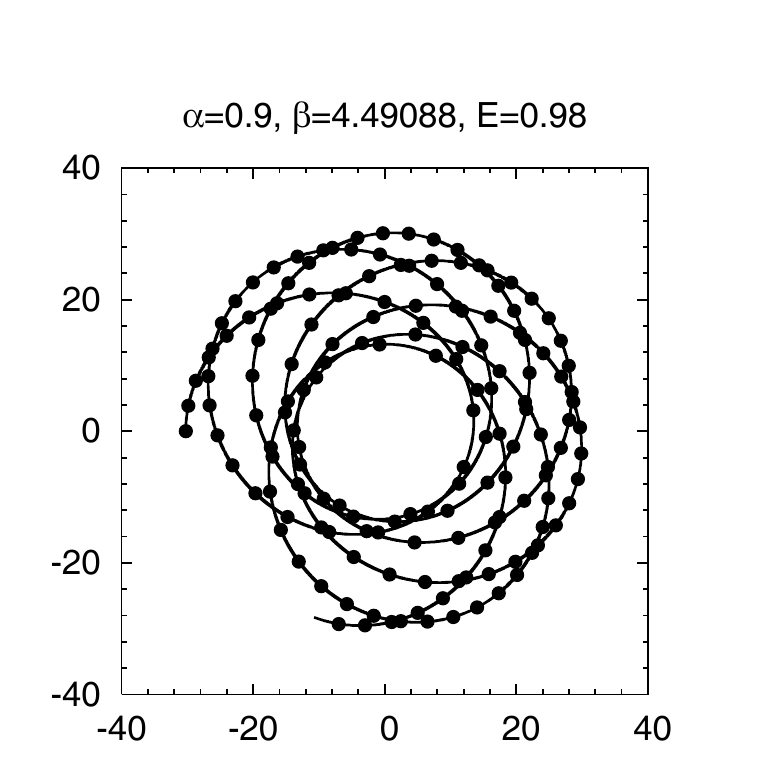}}
\subfigure[][escape orbit, $E=1.87^{1/2}$]{\includegraphics[width=0.3\textwidth]{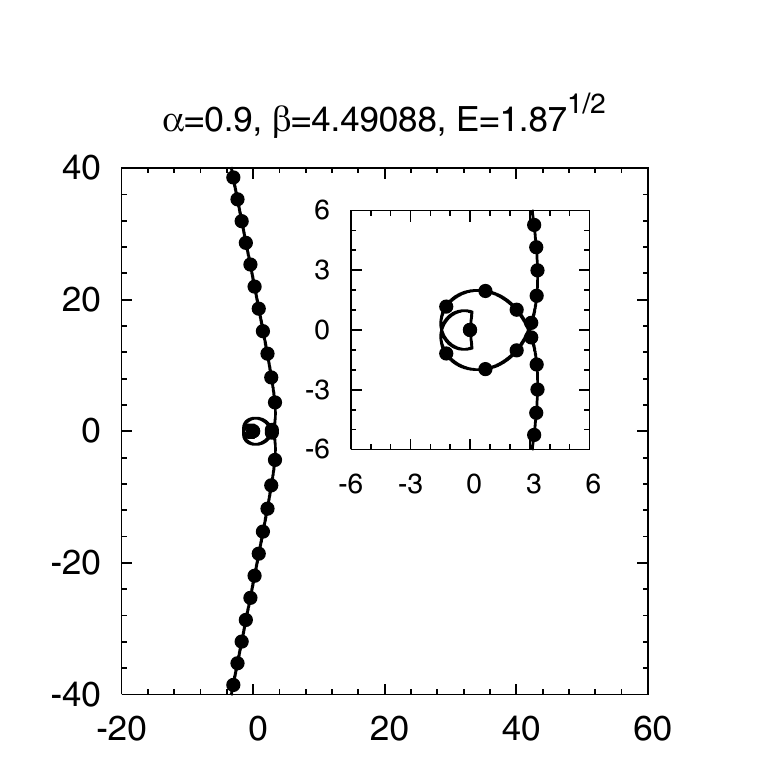}}
\subfigure[][$l$ as function of $\varphi$ for plot (b)]{\includegraphics[width=0.3\textwidth]{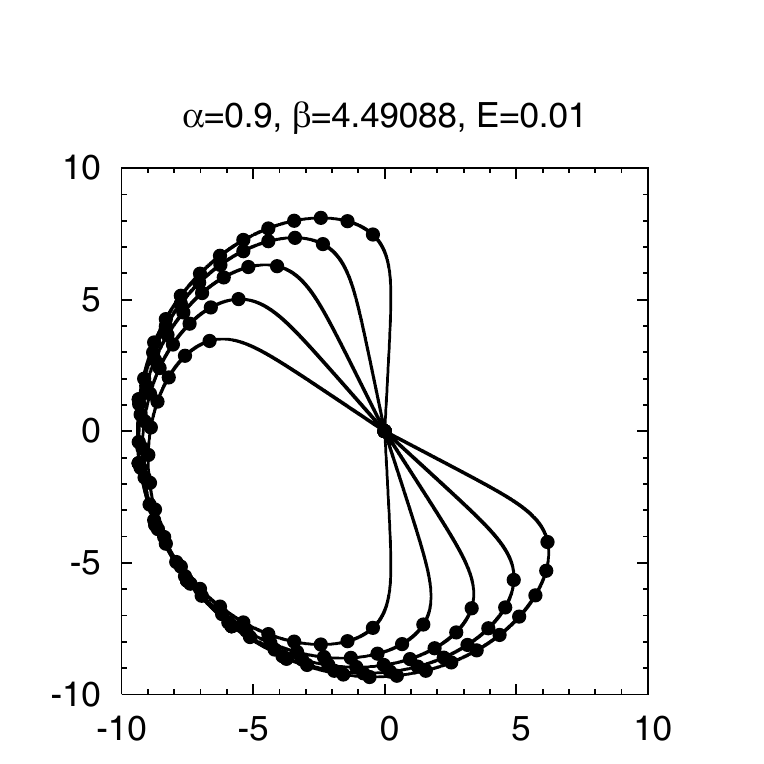}}
\end{center}
\caption{
Particle orbits (a)-(e) $r(\varphi)$ and (f) $l(\varphi)$ for $(\alpha,\beta)=(0.9,4.49088)$.
The dots indicate units of elapsed proper time.
\label{series4b}}
\end{figure}

We exhibit the metric coefficients of this {\sl almost} critical
space--time in Fig.~\ref{series4a}.
The evolution of the monopole space--time towards the critical solution
is similar the one discussed in the previous case,
$(\alpha,\beta) = (1.38585,0)$.
As $\alpha \rightarrow \alpha_{\rm cr}$
the minimum of the metric function $1/g_{rr}$
decreases and reaches zero in the limit.
The true limiting solution then consists of
the non-Abelian interior solution
in the region $0 \le r < r_0$,
and the extremal Reissner--Nordstr\"om
solution with degenerate horizon at $r_0= \alpha$
in the exterior region $r_0 < r < \infty$.

Again, the metric function $-g_{tt}$ becomes increasingly small
in the inner region, $0 \le r < r_0$,
as $\alpha \rightarrow \alpha_{\rm cr}$,
and rises very steeply in the vicinity of $r_0$.
The main difference to the previous case is,
that $-g_{tt}$ is no longer monotonic,
but that it has a minimum close to $r_0$,
before the steep rise.
(Note, that this minimum foreshadows the appearance
of a second minimum in $g_{rr}$ for still larger
values of $\beta$, discussed below.)
In the limit $\alpha \rightarrow \alpha_{\rm cr}$,
$-g_{tt}$ again appears to become singular at $r_0$,
as reflected in Fig.~\ref{series4a} by the Kretschmann scalar.

Considering the motion of particles in the
{\sl almost} critical space--time,
we again select for the angular momentum the value $L=4.3$.
The potential $U$ then again possesses two minima,
giving rise to similar orbits as observed in the previous case.

We display a set of characteristic orbits in Fig.~\ref{series4b}.
At very low energy (a) we observe smooth bound orbits
in the inner region, which are limited by a circle at $r_0$,
that appears to form an outer envelope for the motion.
As the energy increases, the bound orbits in the inner region (b) and (c),
are again hugely distorted,
exhibiting smooth segments alternating with almost straight radial lines.

The energy for the bound orbit in (b) is chosen,
so that the motion is still limited by the circle at $r_0$,
but that the motion now either (almost) proceeds on segments of
that circle or that it proceeds on almost straight radial lines
to and from the center (with some reflection there).

The particle motion is again understood by examining the effective potential.
As before, the abrupt changes in the motion
followed by the straight radial line segments
are caused by the extreme decrease of the effective potential
in the region $r < r_0$, leading to vigorous attraction
for the particle towards the origin.
At the same time the particle needs almost no proper time
to traverse the inner region.

Fig.~\ref{series4a} also exhibits a bound orbit
in the outer region (d), and a scattering state (e).
This scattering orbit reflects the maximum of the potential,
when the particle traverses the inner loops circling the center,
as well as the critical value $r_0$, since the particle
passes also the associated straight radial line segments
close to the center.

\boldmath
\subsubsection{Orbits at $\alpha=0.85$, $\beta=8.282558$}{\label{lastsub}}
\unboldmath

As our last example for motion in the space--time of a
gravitating monopole, we discuss a case with
intermediate Higgs mass,
choosing the parameter set $(\alpha,\beta)=(0.85,8.282558)$.
Here a new phenomenon arises:
the limiting space--time no longer corresponds to
an extremal Reissner-Nordstr\"om solution in the exterior,
but it retains non-Abelian fields there,
which affect the features of the space--time \cite{lue}.

\begin{figure}[p]
\begin{center}
\subfigure[][$-g_{tt}$]{\includegraphics[width=0.4\textwidth]{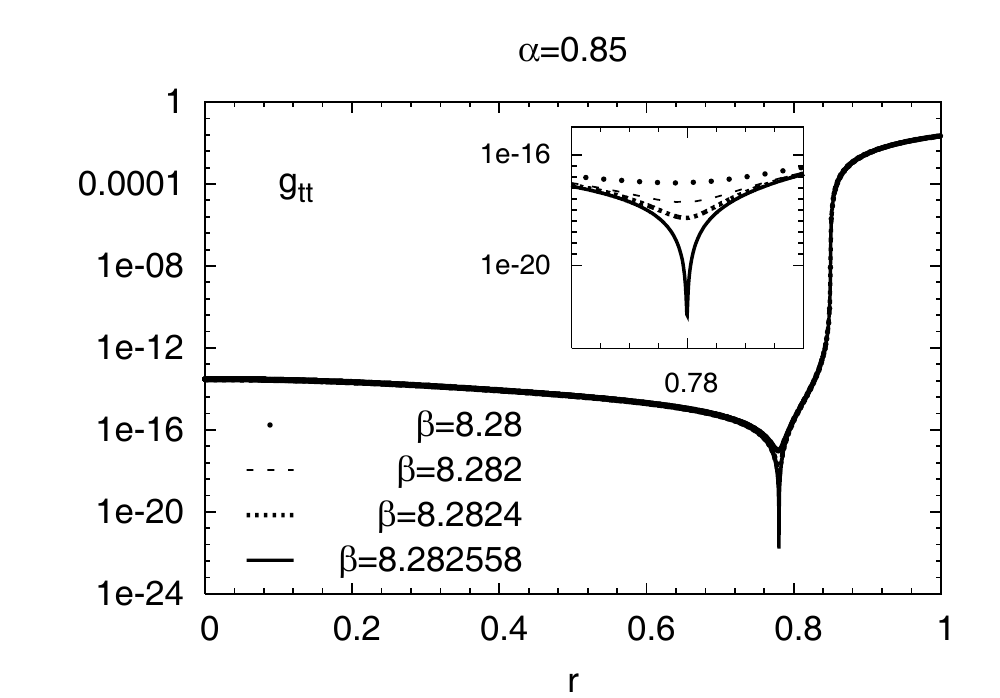}}
\subfigure[][$1/g_{rr}$]{\includegraphics[width=0.4\textwidth]{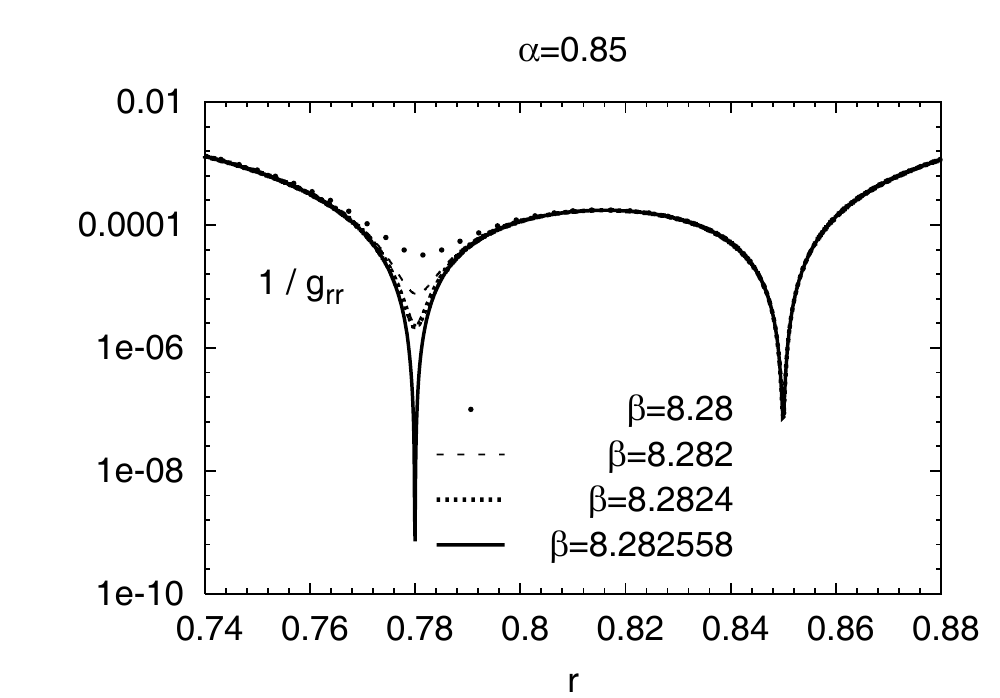}}
\subfigure[][$U$]{\includegraphics[width=0.4\textwidth]{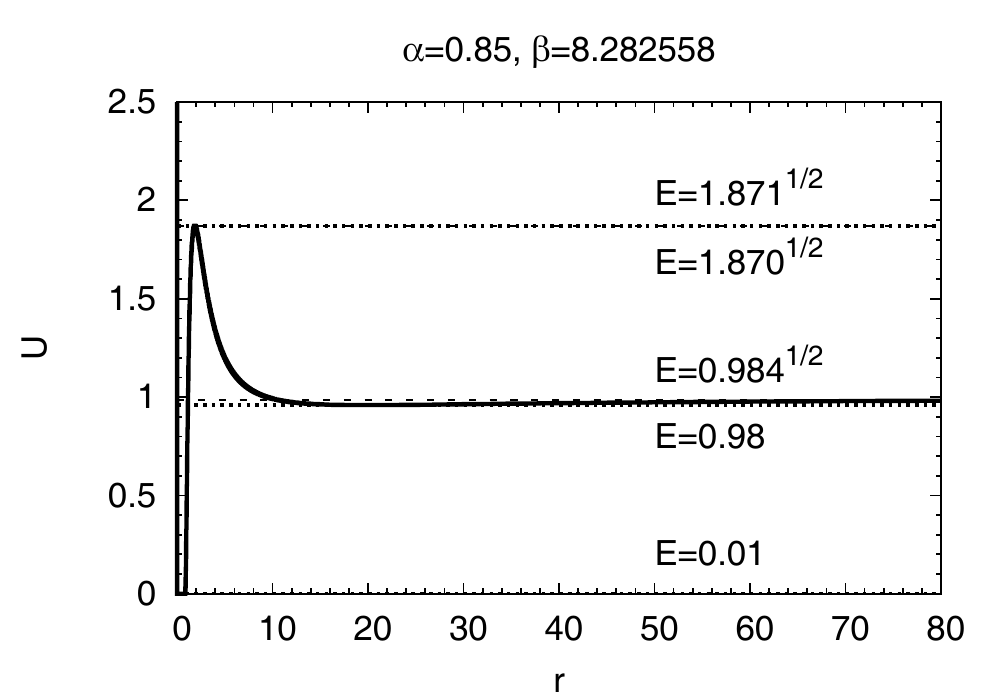}}
\subfigure[][$K$]{\includegraphics[width=0.4\textwidth]{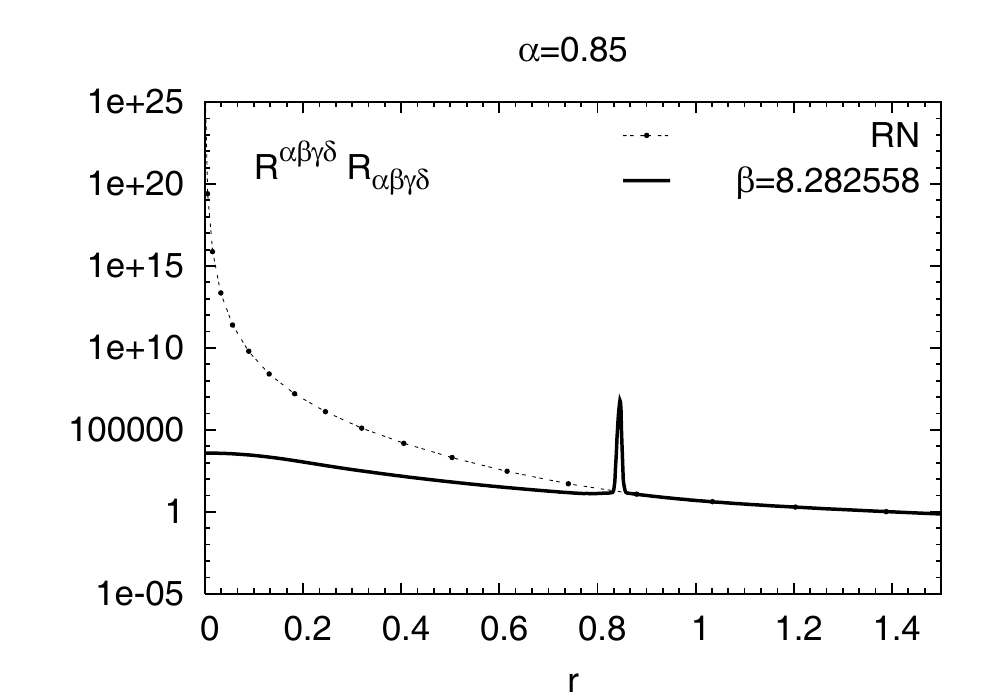}}
\subfigure[][$U_{\rm eff}$]{\includegraphics[width=0.4\textwidth]{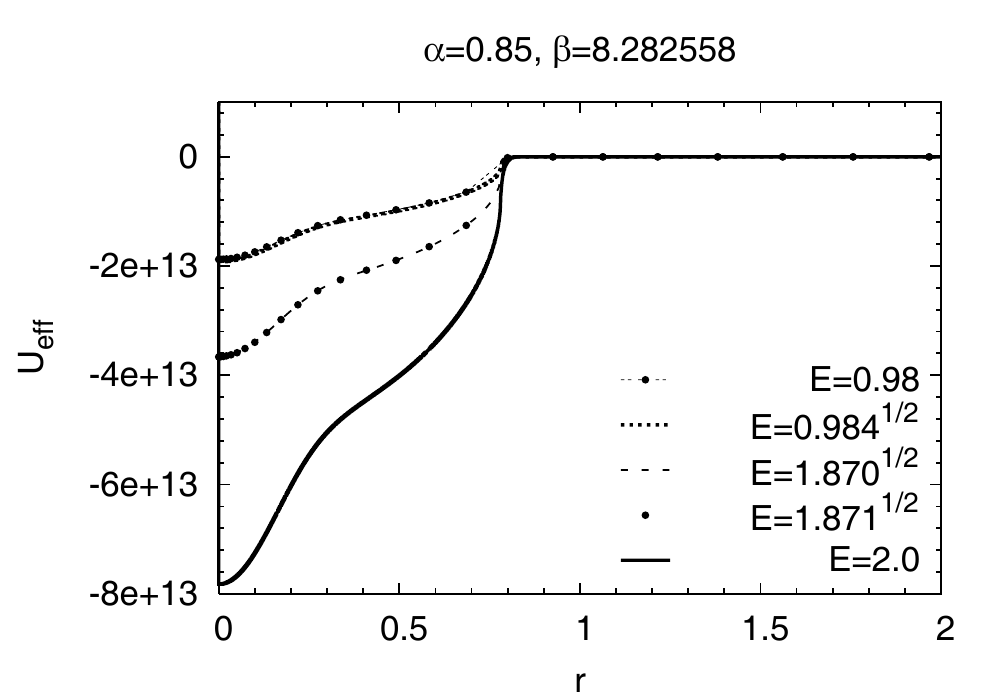}}
\subfigure[][$U_{\rm eff}$]{\includegraphics[width=0.4\textwidth]{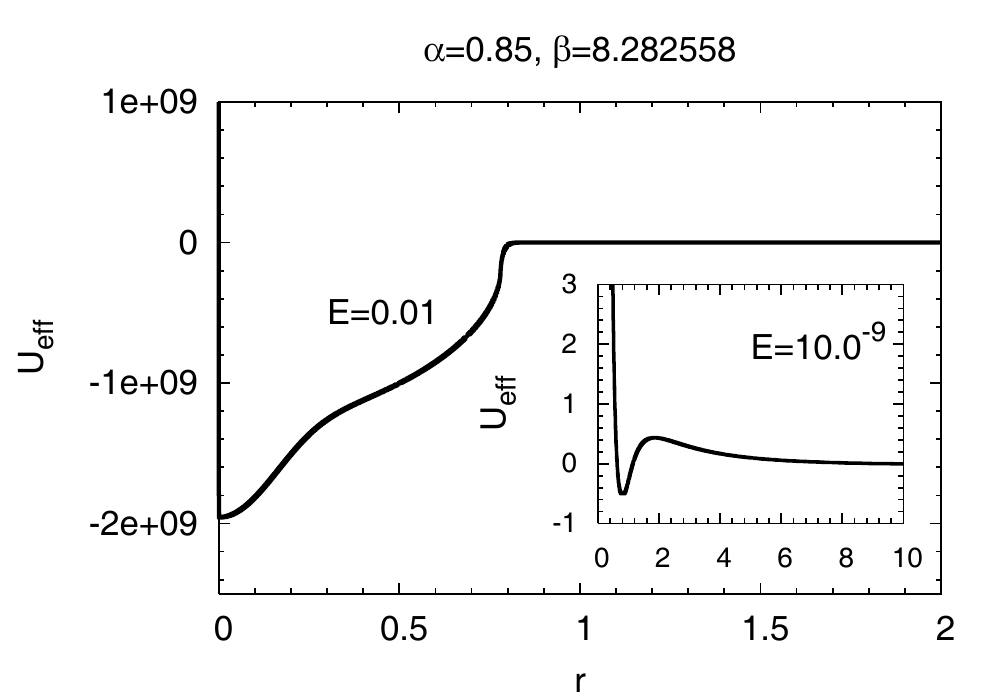}}
\end{center}
\caption{The metric functions $-g_{tt}$ (a) and $1/g_{rr}$ (b),
the potential $U$ (c),
and the curvature invariant $K$ (d),
the effective potential $U_{\rm eff}$ (e) and (f),
versus the radial coordinate $r$
for $(\alpha,\beta)=(0.85,8.282558)$.
(RN denotes the extremal Reissner--Nordstr\"om black hole.)
\label{series5a}}
\end{figure}

\begin{figure}[p]
\begin{center}
\subfigure[][circular orbit, $E=10.0^{-9}$]{\includegraphics[width=0.3\textwidth]{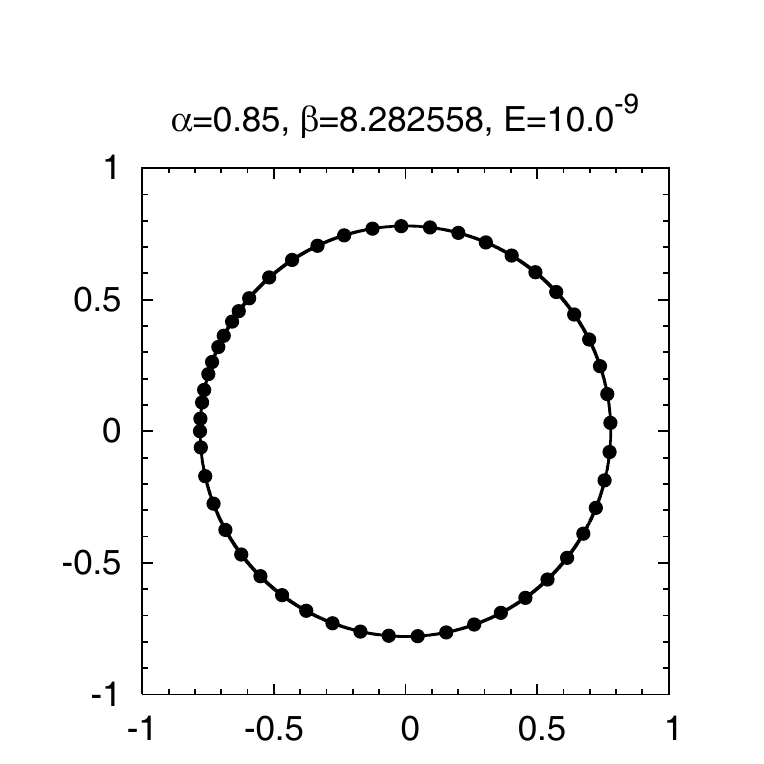}}
\subfigure[][inner region, $E=10.0^{-2}$]{\includegraphics[width=0.3\textwidth]{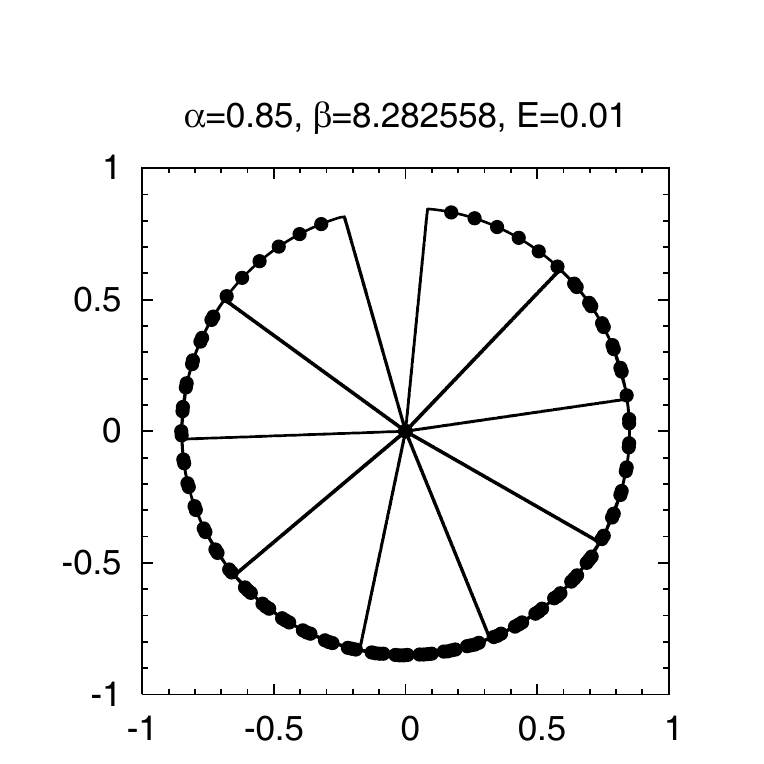}}
\subfigure[][inner region, $E=0.98$]{\includegraphics[width=0.3\textwidth]{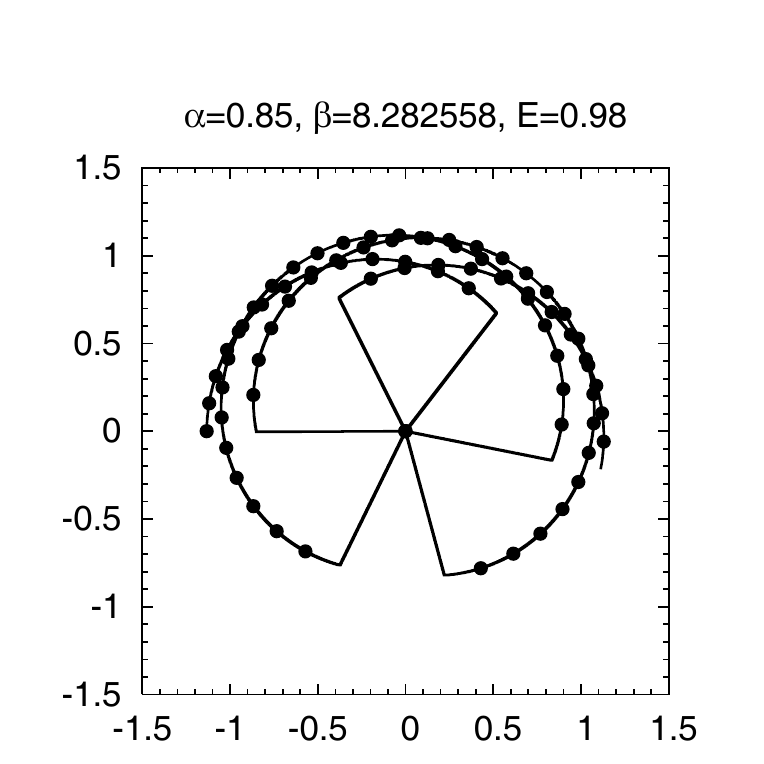}}

\subfigure[][outer region, $E=0.98$]{\includegraphics[width=0.3\textwidth]{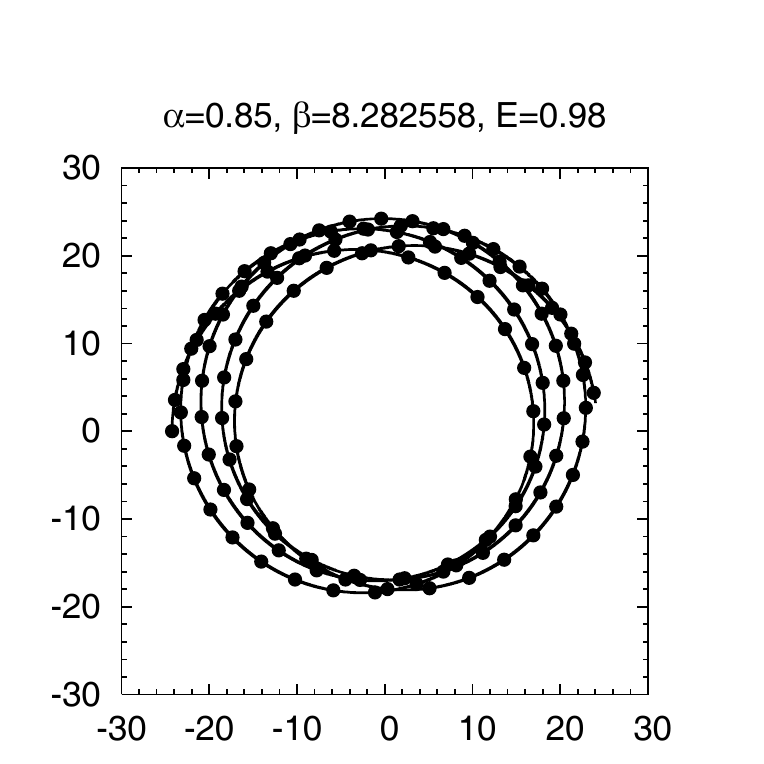}}
\subfigure[][inner  region, $E=0.984^{1/2}$]{\includegraphics[width=0.3\textwidth]{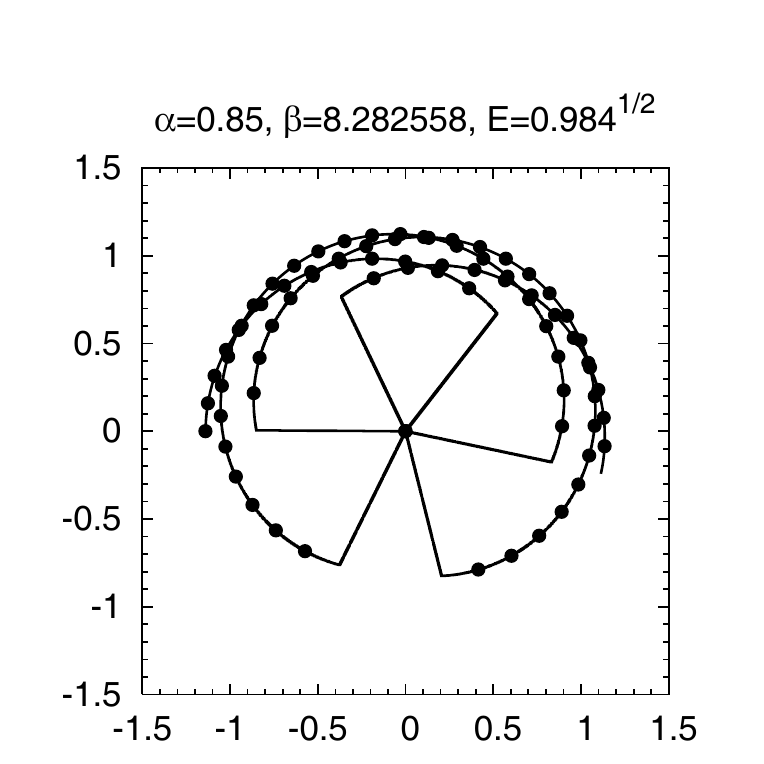}}
\subfigure[][outer region, $E=0.984^{1/2}$]{\includegraphics[width=0.3\textwidth]{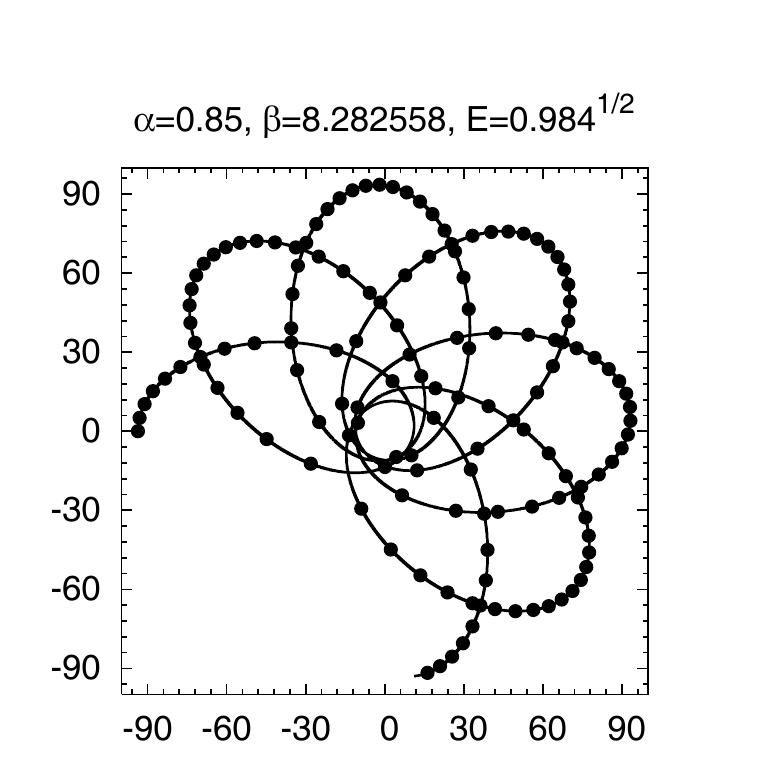}}

\subfigure[][``last'' bound orbit, $E=1.870^{1/2}$]{\includegraphics[width=0.3\textwidth]{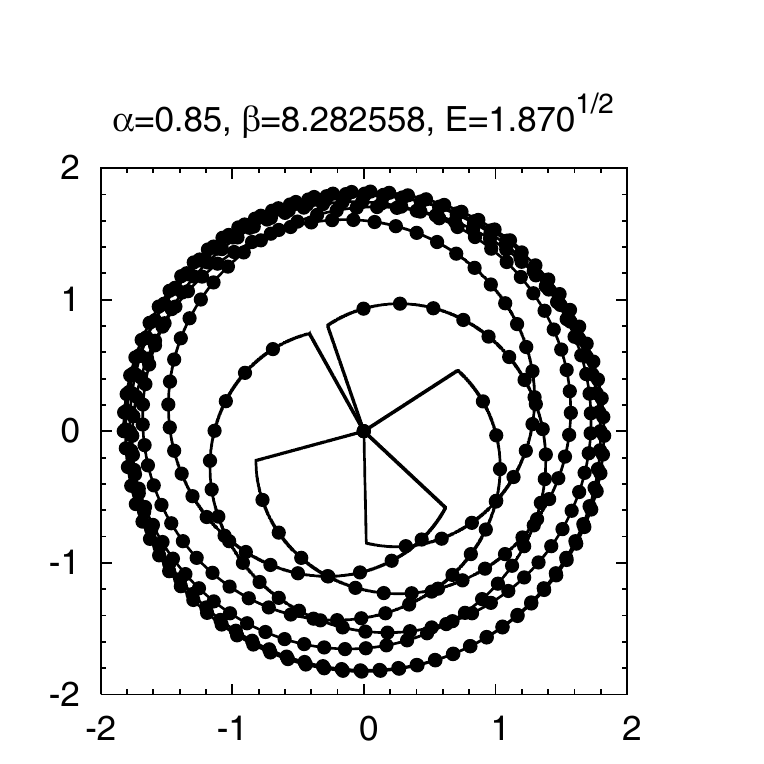}}
\subfigure[][escape orbit, $E=1.871^{1/2}$]{\includegraphics[width=0.3\textwidth]{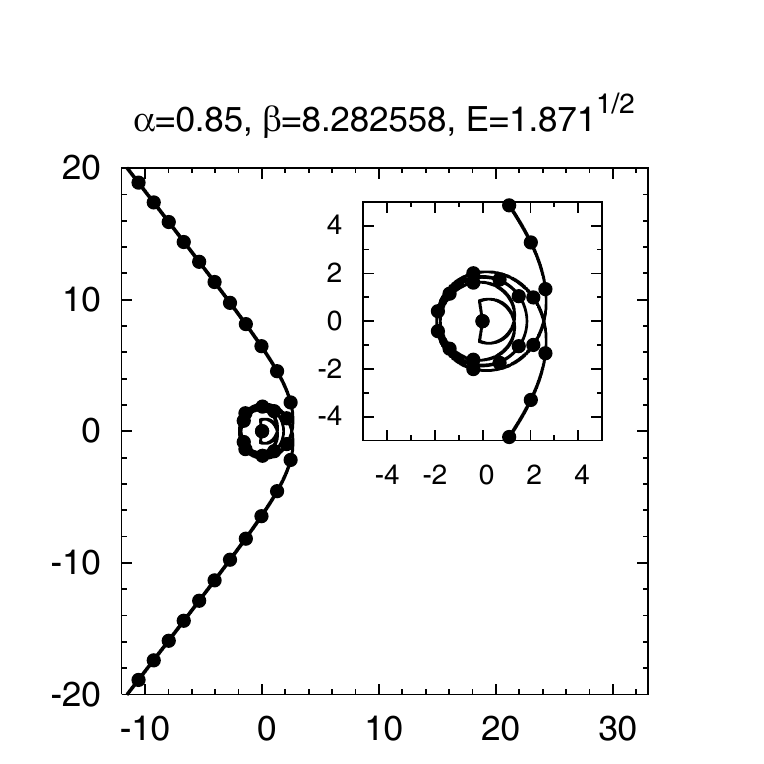}}
\subfigure[][escape orbit, $E=2.0$]{\includegraphics[width=0.3\textwidth]{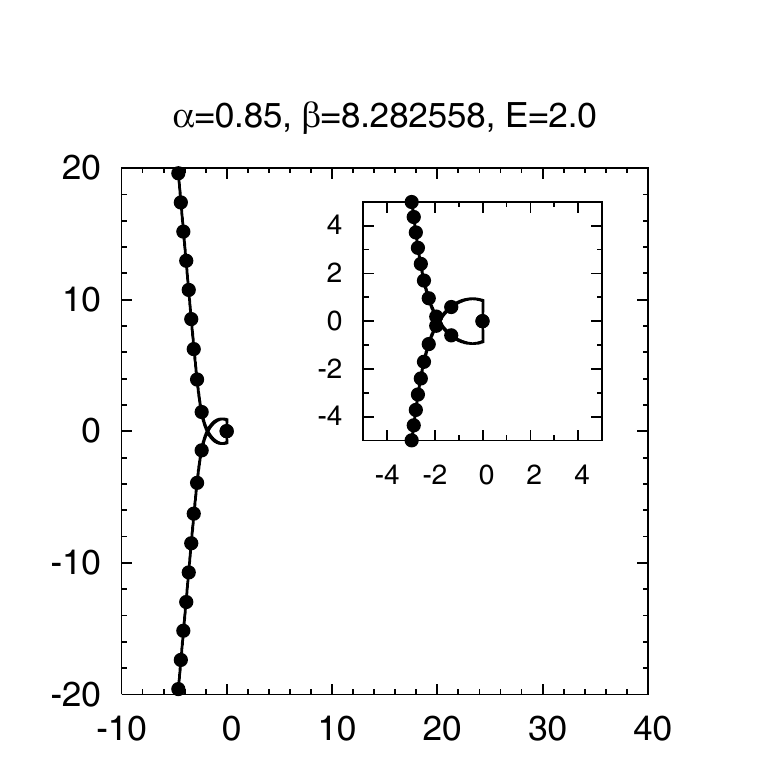}}
\end{center}
\caption{
Particle orbits $r(\varphi)$ for $(\alpha,\beta)=(0.85,8.282558)$.
The dots indicate units of elapsed proper time.
\label{series5b}}
\end{figure}

Let us first consider, how the monopole space--time
evolves in this case as the critical solution is approached.
As in the last example, we fix $\alpha$
and increase $\beta$, until the {\sl almost} critical space--time is reached.
As $\beta \rightarrow \beta_{\rm cr}$
(or analogously as $\alpha \rightarrow \alpha_{\rm cr}$)
the minimum of the metric function $1/g_{rr}$ close to
$r_0=\alpha$ decreases, but it does not reach zero.
Instead, at a certain value of the parameter
a second minimum appears at a value $r^*<r_0$.
As the parameter is further evolved towards its critical value,
it is this new minimum at $r^*$, which reaches zero in the limit,
as anticipated from Fig.~\ref{series5a}.
The old minimum at $r_0$ is no longer affected
by the final approach towards the critical solution.

The true limiting solution thus divides space--time
in a different way and consists of
the interior region $0 \le r \le r^*$
and the exterior region $r^* \le r < \infty$,
where the exterior region may be subdivided into
the intermediate region $r^* \le r < r_0$
and the outer region $r_0 < r < \infty$.
Clearly, the metric of the limiting solution
differs now from the metric of an extremal Reissner--Nordstr\"om black hole
also in the exterior region of the limiting space--time $r^* \le r < \infty$.
Though for $r_0 < r < \infty$ it is still close
to an extremal Reissner--Nordstr\"om space--time
for the chosen parameters.

As in the case above with a single minimum at $r_0$,
the metric function $-g_{tt}$ becomes very small
in the region $0 \le r < r_0$,
when $\beta \rightarrow \beta_{\rm cr}$.
However, unlike that case, it decreases distinctly faster
at $r^*$ than in the overall region $0 \le r < r_0$.
Thus $-g_{tt}$ develops a minimum at $r^*$ quite analogous
to the minimum of $1/g_{rr}$.
As a consequence,
the product of $-g_{tt}$ and $g_{rr}$ changes only slightly
in the vicinity of $r^*$,
while, in contrast, $-g_{tt} g_{rr}$ changes very rapidly
in the vicinity of $r_0$.

The implication for the curvature of the space--time is,
that there appears no singularity at $r^*$.
The Kretschmann scalar remains perfectly smooth there,
as seen in Fig.~\ref{series5a}.
On the other hand, the Kretschmann scalar still becomes
very large at $r_0$, but it will not diverge there in the limit.
The critical space--time is thus expected to
correspond to a space--time with a degenerate
black hole horizon at $r^*$, and with no singularities
at either $r_0$ or at the origin.

Let us now address the orbits of particles and also
of light rays in this {\sl almost} limiting space--time.
We again choose for the angular momentum the previous value
$L = 4.3$. The corresponding potential $U$ is
seen in Fig.~\ref{series5a}.

{\sl Time--like geodesics}

We display a set of characteristic particle orbits in Fig.~\ref{series5b}.
Since $r^*$ is a minimum of $g_{tt}$
the potential $U$ also exhibits a minimum at $r^*$
(the $L^2$ term is negligible here).
For a very small minimal energy
this implies an almost circular motion at $r^*$,
exhibited in (a).
The associated effective potential $U_{\rm eff}$
(Fig.~\ref{series5a} (f)) also assumes its minimal value at $r^*$.

As the energy is increased, the effective potential
changes drastically and rapidly assumes
huge negative values in the region bounded by $r_0$
(and not only the region bounded by $r^*$).
Thus $r_0$ still plays a most significant role for the orbits.
For instance,
the energy for the bound orbit in (b) is again chosen,
such that the motion is limited by the circle at $r_0$.
The motion then proceeds either on segments of
that circle or on almost straight radial lines
to and from the center (with some reflection there).

At somewhat higher energies the bound orbits
in the inner region (c) and (e)
again exhibit smooth segments which alternate with
almost straight radial lines to and from the center,
where the abrupt directional changes occur at $r_0$
(and at the center).
This remarkable role of $r_0$ is preserved also
in the weakly bound orbits (g) and in the escape orbits (h) and (i).
As discussed above, a particle needs very little proper time
to traverse the inner region.

The bound orbits in the outer region (d) and (f), of course,
do not penetrate far enough inside to experience
any influence of $r_0$. These trajectories are ordinary
quasi--elliptic orbits with a perihelion shift.

{\sl Null geodesics}

\begin{figure}[t]
\begin{center}
\subfigure[][$U$]{\includegraphics[width=0.4\textwidth]{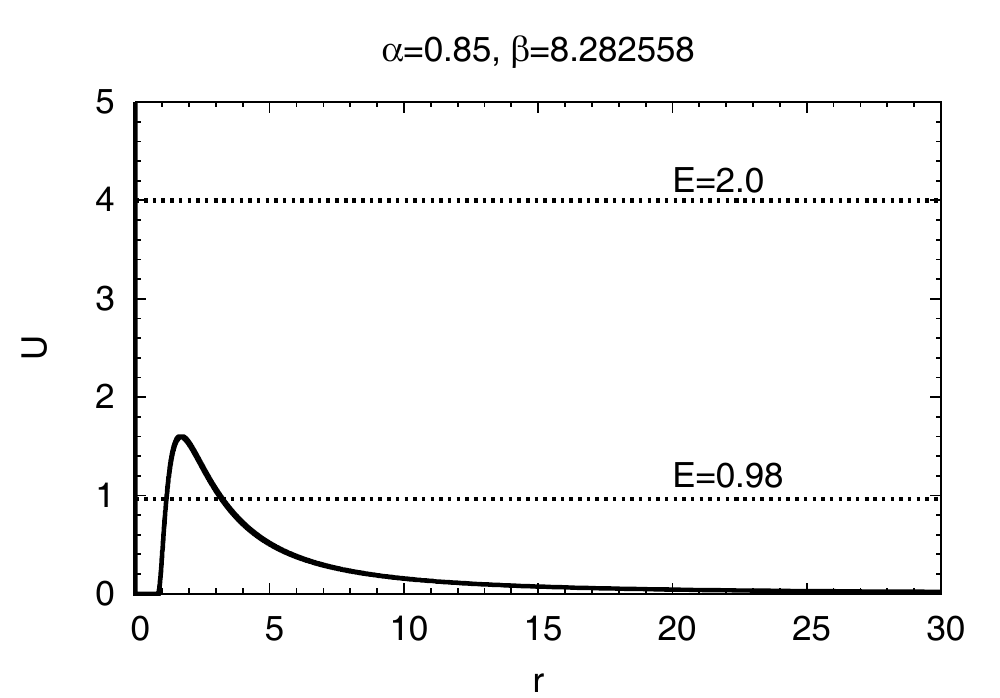}}
\subfigure[][$U$]{\includegraphics[width=0.4\textwidth]{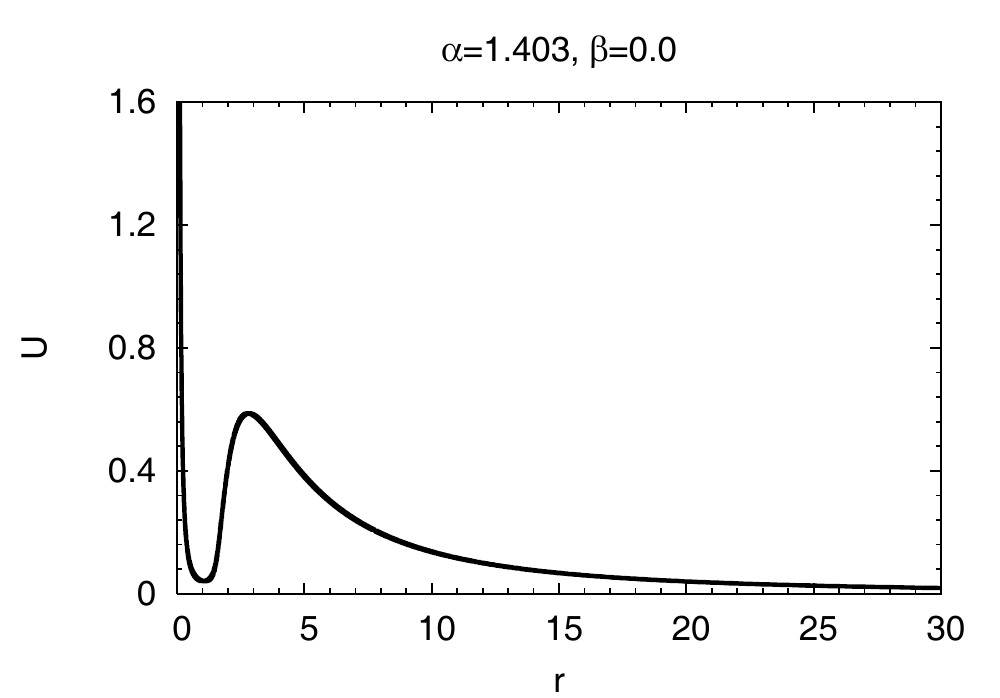}}
\end{center}
\caption{
The potential $U$ for null geodesics
versus the radial coordinate $r$
for $(\alpha,\beta)=(0.85,8.282558)$,
as well as for $(\alpha,\beta)=(1.403,0)$ (subsection \ref{secondsub}).
\label{series5c}}
\end{figure}

\begin{figure}[t]
\begin{center}
\subfigure[][$E=0.98$]{\includegraphics[width=0.3\textwidth]{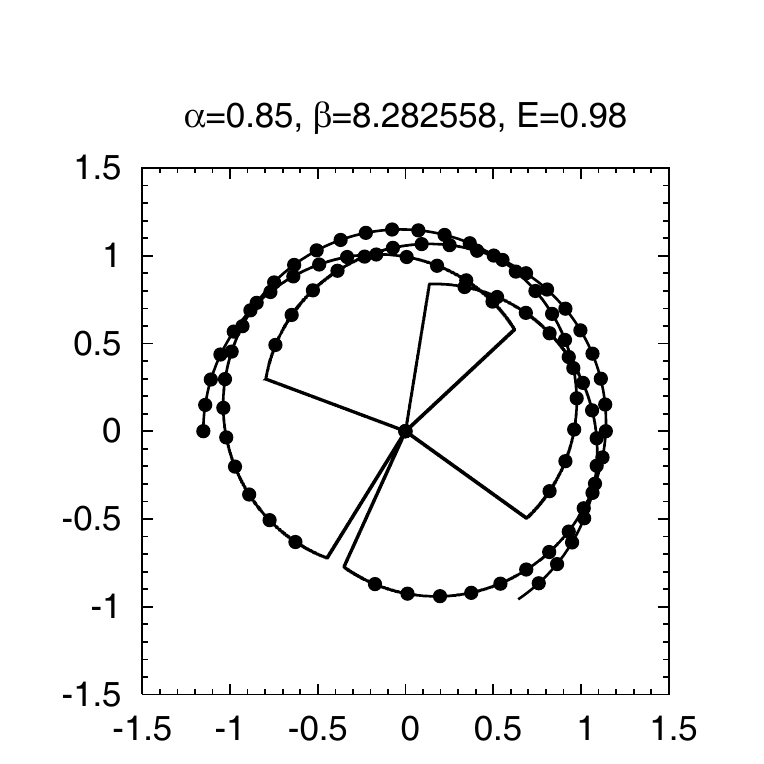}}
\subfigure[][$E=0.98$]{\includegraphics[width=0.3\textwidth]{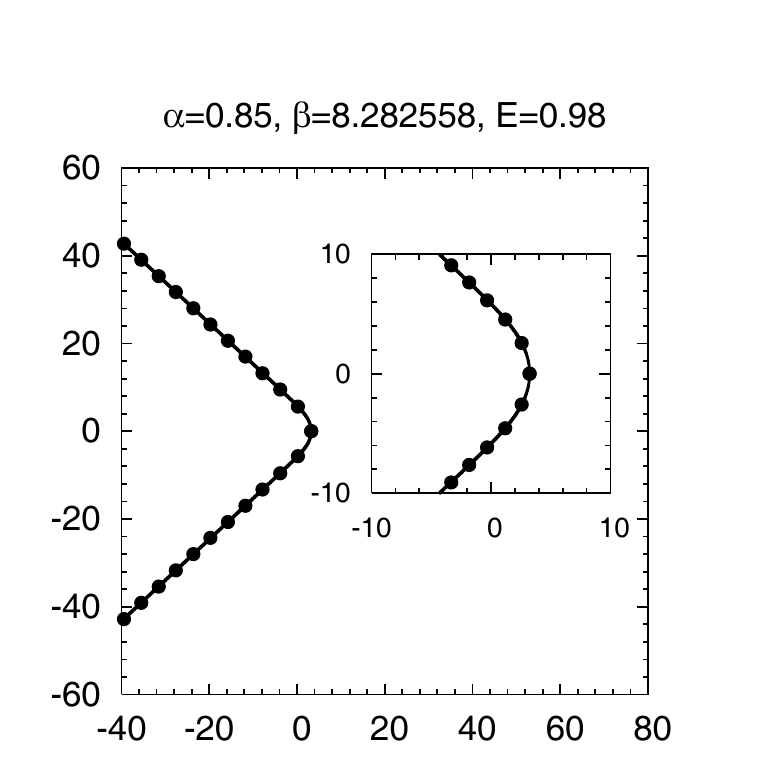}}
\subfigure[][$E=2.0$]{\includegraphics[width=0.3\textwidth]{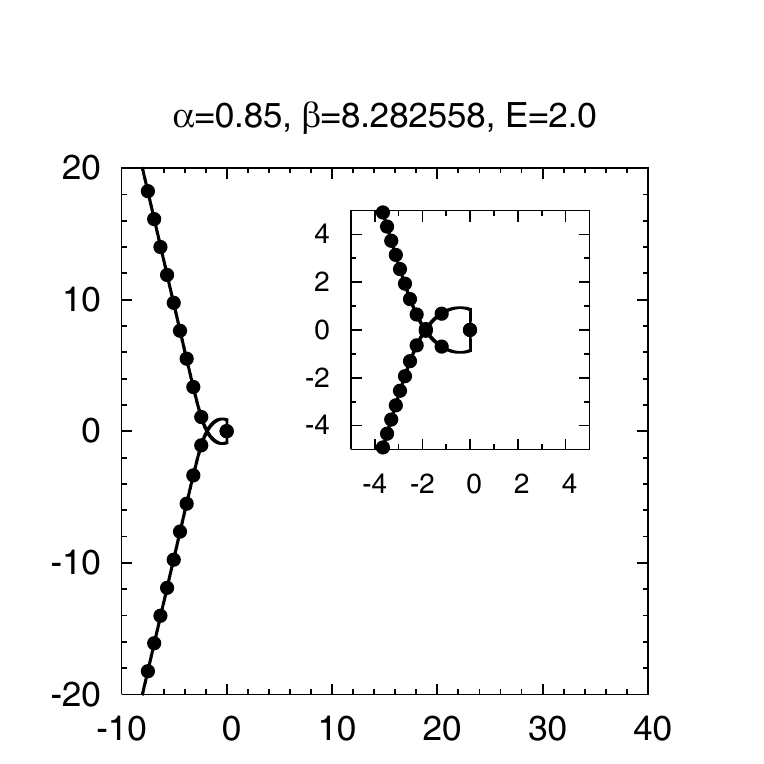}}
\end{center}
\caption{
Null geodesics $r(\varphi)$ for $(\alpha,\beta)=(0.85,8.282558)$.
\label{series5d}}
\end{figure}

We have also considered the null geodesics in this space--time.
The potential $U$ for light rays ($\epsilon = 0$)
is shown in Fig.~\ref{series5c},
while a set of orbits is exhibited in Fig.~\ref{series5d}.
The null geodesics are interesting due to the fact,
that light rays can follow bound geodesics,
as evident from the form of the potential $U$.
Thus light can be captured by the source,
but does not fall into any singularity,
since the space--time is regular.

The shape of the null geodesics is similar
to the shape of the time--like geodesics, as seen in Fig.~\ref{series5d}.
The bound orbits (a)
again exhibit smooth segments which alternate with
almost straight radial lines to and from the center,
with abrupt directional changes at $r_0$
(and at the center),
related to the fact
that the effective potential becomes very large and negative
in the region within the radius $r_0$.
This peculiar effect on the motion
at and within radius $r_0$ is also seen in the scattering orbits
at high enough energy (c).
The scattering orbits at lower energy (b)
are deflected before they can experience
any influence of $r_0$.

\section{Conclusions and outlook}\label{Sec:conclusion}

An accurate analysis and true interpretation
of a given gravitational field can only be obtained
through the exploration of the geodesics
of particles and light rays in this space--time.
Such an investigation is particularly important in regions,
where the space--time metric exhibits unusual behavior,
as for example in the vicinity of a black hole event horizon.

In this paper we have discussed the motion of particles and light rays
in the gravitational field of a magnetic monopole
which is characterized by two dimensionless constants, $\alpha$ and $\beta$,
where $\alpha$ signifies the relative strength of gravity,
while $\beta$ is the ratio of the Higgs boson to the vector boson mass
in the non-Abelian gauge theory.

The space--time of a non-Abelian monopole is globally regular.
Its curvature invariants are finite everywhere.
However,
globally regular monopoles cannot exist for large values of
the gravitational coupling strength.
Simple dimensional reasoning shows, that black holes should form,
when $\alpha$ reaches values on the order of one \cite{gmono}.
Indeed, at critical values of the parameters
($\alpha_{\rm cr},\beta_{\rm cr}$)
the space--time changes dramatically
and the metric becomes singular.

As the critical space--time is approached
for small values of $\beta$,
the metric coefficient $1/g_{rr}$ tends to zero at $r_0=\alpha_{\rm cr}$,
and the metric coefficient $g_{tt}$
tends to zero in the interval $0<r<r_0$.
The effects on the particle orbits are then astounding.
The orbits traversing the region, where $r_0$ is located,
are hugely distorted as compared to ordinary bound orbits.
While the orbits evolve smoothly until they reach the vicinity of $r_0$,
they then change abruptly direction
and approach the center in almost straight radial lines.
There they are reflected,
and move in almost straight radial lines out to $r_0$,
where they change abruptly direction again and evolve smoothly further.
This intriguing pattern of movement is caused by the
steep drop of the effective potential for $r < r_0$,
leading to a vigorous attraction of particles
or light rays in this region.
Also, particles need almost
no proper time to traverse this region ($r < r_0$).

For larger values of $\beta$, the critical space--time is
distinctly different \cite{lue}. The metric coefficient
$1/g_{rr}$ acquires a second minimum at $r^*<r_0$,
which tends to zero in the limit,
while the minimum at $r_0$ remains finite.
Still it is $r_0$, and the steep drop of the effective potential
associated with it,
which strongly dominates the orbits in the interior,
causing abrupt directional changes and vigorous attraction
towards the center, along with almost no lapse of proper time.

We note, that this kind of research may be extended
to metrics associated with stationary axially symmetric solutions
of the Einstein field equations.
There are interesting examples where counterrotating horizons appear
\cite{counter}
or negative horizon masses
\cite{mass}.
Such unusual features of various space--times
may be explored best
through the study of the motion of particles and light.
{This may be extended to the study of particles with spin
\cite{ChiconeMashhoonPunsly05}.}

Another extension of the present considerations is the study
of the motion of satellites and stars
in the gravitational field of a modified gravitational theory.
This may have applications to the Pioneer anomaly, the flyby anomaly,
or the increase of the astronomical unit
\cite{LaemmerzahlPreussDittus06},
which are all problems which are unresolved within standard general relativity.


\subsection*{Acknowledgement}

We would like to thank P. Breitenlohner, B. Kleihaus, and D. Maison
for valuable discussions.
V.K. thanks the German Academic Exchange Service DAAD
and C.L. the German Aerospace Center DLR for financial support.

\labelsep20pt

\end{document}